\documentclass[useAMS,usenatbib]{mn2e}

\usepackage{graphicx}

\usepackage{natbib}
\usepackage{bm}

\newcommand{\be}{\begin{equation}}
\newcommand{\ee}{\end{equation}}

\newcommand{\ba}{\begin{eqnarray}}
\newcommand{\ea}{\end{eqnarray}}

\newcommand\eg{\textit{e.g.}}
\newcommand\cf{\textit{cf.}}

\newcommand{\Bf}{{magnetic field}}
\newcommand{\Bfs}{{magnetic fields}}

\title[Structure of magnetic fields in intracluster cavities]{Structure of magnetic fields in intracluster cavities}

\author[]{Konstantinos Nektarios Gourgouliatos$^1$, Jonathan Braithwaite$^2$ and Maxim Lyutikov$^1$ \\
$^1$ Department of Physics, Purdue University, \\
 525 Northwestern Avenue,
West Lafayette, IN
47907-2036 \\
$^2$ Argelander Institut f\"ur Astronomie, University of Bonn,\\
Auf dem H\"ugel 71, D-53121 Bonn } 

\begin{document}

\date{Accepted -. Received -; in original form -}
\pagerange{\pageref{firstpage}--\pageref{lastpage}} \pubyear{-}
\maketitle

\label{firstpage}

\begin{abstract}
Observations of clusters of galaxies show  ubiquitous presence of  X-ray cavities, presumably blown by the AGN jets. We consider magnetic field structures of these cavities. Stability requires that they contain both toroidal and poloidal magnetic fields, while realistic configurations should have vanishing magnetic field on the boundary. For axisymmetric configurations embedded in unmagnetized plasma, the  continuity of poloidal and toroidal magnetic field components on the surface of the bubble then requires solving the elliptical Grad-Shafranov equation with both Dirichlet and Neumann boundary conditions. This leads to a double eigenvalue problem, relating the pressure gradients and the toroidal magnetic field to the radius of the bubble. We have found fully analytical stable solutions. This result is confirmed by numerical simulation. We  present synthetic X-ray images and synchrotron emission profiles  and evaluate the rotation measure for radiation traversing the bubble. 
\end{abstract}


\maketitle

\section{Introduction}

X-ray and radio  observations of intracluster medium (ICM)  demonstrate the presence of cavities  produced by the AGN jets \citep{McNamaraNulsen, Fabian2000, Bir2004, D2008, D2010}. Cavities are filled with jet-supplied plasma  and \Bfs\ and are  observed both as depressions in X-ray surface brightness and through intrinsic synchrotron emission.  The nature of the plasma inside the cavity is not determined yet, a way to determine whether the plasma is thermal or not is through Sunyaev-Zel'dovich effect \citep{PES2005}. Most cavities are not expanding supersonically and, thus, are  in pressure balance  with  their surroundings. Their rise in the cluster potential is caused by buoyancy.

Cavities are long lived entities, surviving on time scales much longer than the time scale of buoyant rise.  On the other hand, fluid (non-magnetic) simulations show that light  bubbles are disrupted on one scale height \citep{Robinson04,Churazov01,Bruggen02,JonesDeyoung}, but there have also been works which suggest that purely hydrodynamics effects stabilise the cavities \citep{RM2005,SB2010}. Magnetic field can stabilize the  cavities via two mechanisms. First, \Bf\ in the ICM drapes around expanding cavities \citep{Lyutikovdraping} forming a layer of nearly equipartition, which stabilizes the bubble against Kelvin-Helmholtz and Rayleigh-Taylor instability \citep{DursiBubble,DursiPfrommer}. Secondly, if the AGN jet which blew the bubble carries large-scale \Bf, this will stabilize the bubble.

The presence of large scale \Bfs\ in jets is a natural consequence of magnetic launching. It is generally accepted that AGN jets are produced (accelerated and collimated) by large scale electromagnetic forces originating  either near the central black hole \citep[\eg~][]{BlandfordZnajek} or above the  accretion disk \citep{BlandfordPayne,lws87}. It is expected that after the end of the active jet phase, the \Bf\ in the cavity will settle down to a stable equilibrium. In this paper we discuss the properties of such a state. Toroidal fields are dominant in jets after expansion takes place, however a purely poloidal or toroidal field is unstable; therefore it will relax to a structure which contains both components. Thus, the problem of ICM bubbles can be split into two parts, the first one is the structure of a stable cavity; the second one is the behaviour of this cavity as it rises. In this paper we are studying the structure of a static cavity after the relaxation has taken place.

 As a model problem we are looking for an axisymmetric structure of \Bf\ within a spherical cavity. Axisymmetric equilibria
are not only simpler to construct analytically, but there is also good reason to believe that they are the simplest and most fundamental configurations, as the effect of rotation leads to an axially symmetric sturcture \citep{L2006}. In the case of fluid stars,  very general arguments prove that both purely poloidal and purely toroidal \Bfs\ are unstable \citep{Tayler73,FlowersRuderman}. As a result, stable equilibria  must have considerable toroidal \Bf\ component  \citep[\cf][]{Prendergast}. Using numerical methods, the relaxation of an arbitrary magnetic field in a star into an axisymmetric equilibrium has been studied (\citealt{BraithwaiteNordlund2006,Braithwaite2009}; see also \citealt{Lyutikov2010}), and recently \citet{Braithwaite2010} has used similar methods to study the formation of equilibria in X-ray cavities, finding that an arbitrary magnetic field inside a bubble can relax into an axisymmetric equilibrium with poloidal and toroidal components of roughly equal strength. A rising bubble shall experience a stronger stagnation pressure at the top and a weaker pressure at the side leading to deformation and probably flattening of the bubble. 

Here, in particular, we are looking for solutions with vanishing  total \Bf\  on the surface of the bubble. This requirement eliminates unphysical surface currents. This  excludes, in particular, analytical force-free solutions, like the spheromak solution. Spheromaks are solutions of force-free magnetic fields in axial symmetry \citep{CK1957}. They have been invoked before in the context of magnetic bubbles \citep{Tang2008}. However, these solutions have one particular drawback that excludes them as reasonable solution for \Bf\ structure in ICM cavities: they cannot have both poloidal components equal to zero on the surface. Thus, connecting them to unmagnetized plasma outside requires surface currents.  For example, for the basic spheromak solution the surface current is $\propto \sin \theta$. In addition to that, the pressure on the surface of a spheromak varies from zero on the axis and goes to a maximum value on the equator, embedding this structure in a medium with constant pressure will lead to its deformation. It is possible that the system achieves equilibrium after the deformation, this problem has been treated numerically by \cite{Braithwaite2010}.

Since spheromak solutions correspond to linear force-free fields, a large number of high multipole solutions can be chosen to construct a system with nearly isotropic surface pressure. Intracluster cavities are blown by AGN jets which may carry large scale magnetic fields, nearly all models of AGN jet launching and acceleration require the presence of large  scale magnetic fields near the central black hole \citep{BlandfordZnajek,GMT2003}; we assume that large scale fields survive up to the large distances. The jet carries mostly toroidal magnetic field; after the ending of an  intermittent  AGN activity, the magnetic field structure within the bubble should reach a stationary state approximately on one Alfv\'en crossing time.  We assume that during this relaxation the magnetic  field within the bubble preserves a large scale component while small scale fields will behave similarly to an additional pressure  component. In this paper we discuss the structure of magnetic field in these relaxed bubbles.  

The magnetic energy in the bubble must be confined by the gas, whose pressure must therefore be lower inside the bubble than outside. Either the Lorentz force and pressure gradient force balance each other throughout the volume of the bubble, or the forces act only at the surface of the bubble where a singularity in the Lorentz force (a current sheet) is balanced by a discontinuity in the gas pressure, in which case the interior of the bubble can be force-free. Whilst the latter is necessary in the case where the magnetic field dominates over the gas pressure in the interior, i.e. as plasma-$\beta\rightarrow 0$, it seems physically unlikely in the high-$\beta$ case -- therefore we consider here non-force free solutions with no current sheet at the boundary. The confinement in this type of solutions is not localized on the surface of the system as it has been discussed for the case of jets \citep{S2010} but it is extended in the whole volume of the system. 

Other options we have attempted, like a purely poloidal field with some pressure, in addition to the stability issues, require surface currents. Our solution has a magnetic field with both poloidal and toroidal components and also a plasma pressure. We describe the solution and its physical meaning in the following section.

 \section{Magnetic field in the cavity}

\subsection{The Grad-Shafranov equation}

In   MHD equilibria the  Lorentz  force is balanced by pressure gradient; we shall assume that gravity is not important as we constrain our interest to magnetic cavities and not to self-gravitating systems such as magnetic stars, so the equilibrium condition is
\be
\bm{J} \times \bm{B} = \nabla p\,.
\label{MHD}
\ee
Axially symmetric magnetic  fields can be written as
\be
\bm{B}={  \nabla P \times {\bf \hat{e}}_\phi  + 2 I {\bf \hat{e}}_\phi \over r \sin \theta }  ,
\label{B0}
\ee
where $2\pi P$ is the poloidal magnetic flux and $ cI$ is the poloidal current enclosed
by an axially symmetric loop  \cite[\eg][]{Shafranov}; we set speed of light to unity. Stationary solutions require $I=I(P)$,
 and 
 the force balance  (\ref{MHD})  gives the Grad-Shafranov  equation
\be 
 \Delta ^\ast P  +F(P) r^2 \sin ^2\theta + G(P)=0
 \label{GS}
 \ee
 where
\be
\Delta ^\ast  = {\partial^2\over\partial r^2} + {\sin \theta \over r^2}  {\partial\over\partial\theta} \left({1\over \sin \theta}{\partial\over\partial\theta}\right)
\ee
is the Grad-Shafranov operator, $ G(P)= 4 I I' $, 
 $I=I(P)$ and $F=4\pi {\rm d} p /{\rm d} P$ are functions of $P$. 
 
\subsection{Solution}

We are looking for solutions of (\ref{GS}) with vanishing \Bf\ on the surface, as we have assumed the that outside plasma is unmagnetized. 
This requirement leads to  an unusual mathematical problem: we seek solutions of an elliptical equation (\ref{GS}) of a function $P$  while matching both its value and its derivative on the boundary. 

In other words, on the boundary  both the flux function $P$ and its  normal derivative $\partial_r P$ should be continuous. 
This makes an elliptical equation (\ref{GS}), formally,  over-constrained, as the solution should satisfy both Neumann and Dirichlet boundary conditions.

On the other hand, the  Grad-Shafranov equation (\ref{GS}) has two unknown function $F$ and $G$, which, as we show below, can be chosen in such a way to satisfy the over-constraining boundary condition. This calls for an unusual mathematical problem, to find the equation of the type  (\ref{GS}), where $F(P)$ and $G(P)$ are functions of the solution only, so that the solution itself has to satisfy both Neumann and Dirichlet boundary conditions. 

We have found that a linear dependence of the pressure and toroidal field function $I$ on $P$, lead to solutions of the desired properties, thus we choose $F(P)=F_{0}$ and $G(P)=\alpha^{2} P$. The angular part of the differential operator of the Grad-Shafranov equation admits eigenfunctions which are expressed in terms of Legendre polynomials in the following way
\be
E_{1}=\sin^{2}\theta \frac{{\rm d} P_{l}(\cos\theta)}{{\rm d} (\cos \theta)}\,.
\ee
The fact that the pressure term is multiplied by $\sin^{2}\theta$ constrains our solutions to dipole solutions, any higher multipole leads to imbalance between the Lorentz and the pressure forces, for instance see appendix A of \cite{GV2010}. Following the method of separation of variables for $P=\sin^{2}\theta f(r)$ the $\sin$ terms drop out and we are left with an ordinary differential equation for $f(r)$ 
\be
f''-\frac{2}{r^2}f+F_{0}r^{2}+\alpha^{2}f=0\,.
\ee
This equation admits analytical solutions
\begin{eqnarray}
f=c_{1}\Big(\alpha \cos(\alpha r)-\frac{\sin(\alpha r)}{r}\Big) \nonumber \\
 +c_{2}\Big(\alpha \sin(\alpha r)+\frac{\cos(\alpha r)}{r}\Big)-\frac{F_{0}}{\alpha^{2}}r^2  \,.
\end{eqnarray}
By demanding that the flux is not infinite at the origin we set $c_{2}=0$, and then we observe that the solution we have is a spheromak superposed with a uniformly twisted magnetic field.
\begin{eqnarray}
P=\sin^{2}\theta\left[c_{1}\Big(\alpha \cos(\alpha r)-\frac{\sin(\alpha r)}{r}\Big)-\frac{F_{0}}{\alpha^{2}}r^2\right]  \,.
\label{Solution}
\end{eqnarray}
\begin{figure}
\includegraphics[width=0.8\linewidth]{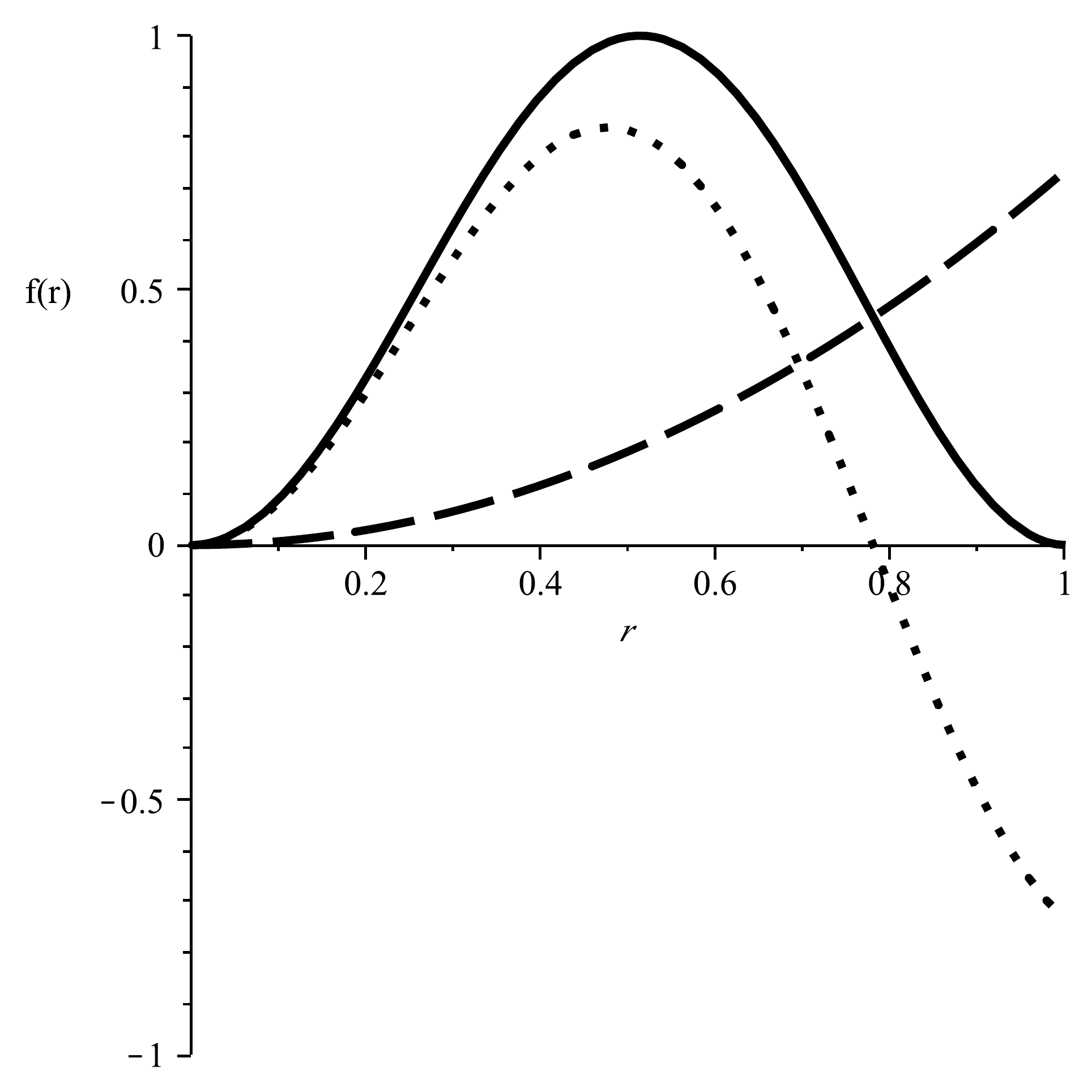}
\caption{Plot of the solution as it appears in expression~(\ref{Solution}) (solid line), the spheromak component of the field (dotted line) and the twisted field (dashed line).}
\label{Decomp}
\end{figure}
The next step is to enforce the boundary conditions. The flux is confined within a sphere of unit radius $r_{0}=1$, this translates to $f(1)=0$. We determine $F_{0}$
\begin{eqnarray}
F_{0}=c_{1}\alpha^{2}(\alpha \cos \alpha -\sin \alpha)\,.
\end{eqnarray}
Then we impose the condition that there are no surface currents, which is $f'(1)=0$, so $\alpha$ is given by
\begin{eqnarray}
\tan \alpha=\frac{3 \alpha}{3-\alpha^2}\,.
\end{eqnarray}
The smallest positive root we find is $\alpha=5.76$ and it is independent of the choice of normalization, for this root the $f(r)$ is zero at $r=0$ and $r=1$ and it is non-zero in the interval between, see Fig.~1; had we chosen a greater root for $\alpha$ we would have taken a solution where the flux becomes zero within the cavity as well being zero at $r=0$ and $r=1$. For this root of $\alpha$ we find that $F_{0}=182.67c_{1}$.
The explicit form for the field is the following:
\begin{eqnarray}
B_{r}=\frac{2 \cos \theta}{r^2}f(r)\,, \\
B_{\theta}=-\frac{\sin \theta}{r} f'(r)\,, \\
B_{\phi}=\frac{\alpha \sin \theta}{r}f(r)\,,
\end{eqnarray}
where the function $f(r)$ and $\alpha$ are determined as we have described above, Fig.~(\ref{Bcomponents}). 
\begin{figure}
\includegraphics[width=4.0cm]{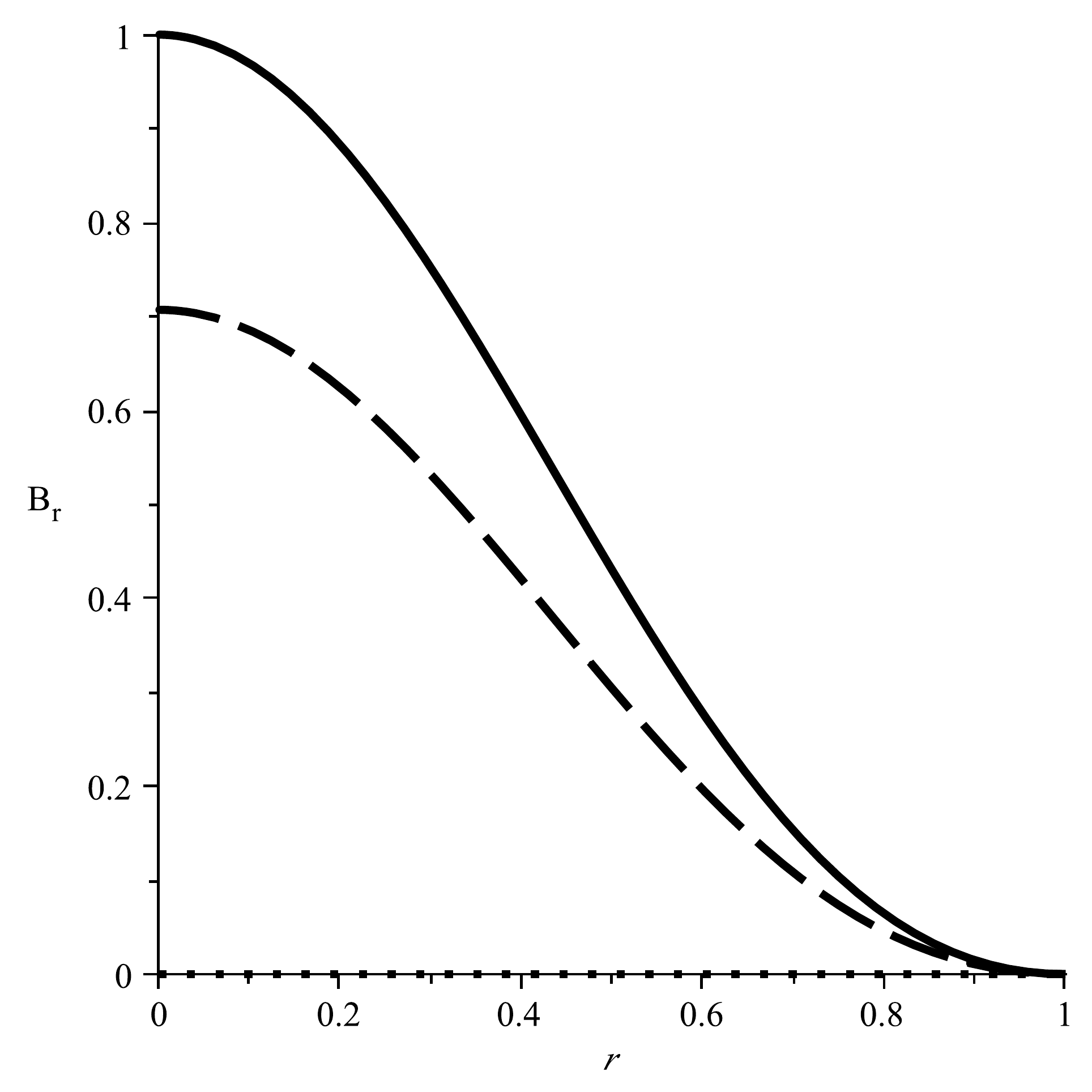}
\includegraphics[width=4.0cm]{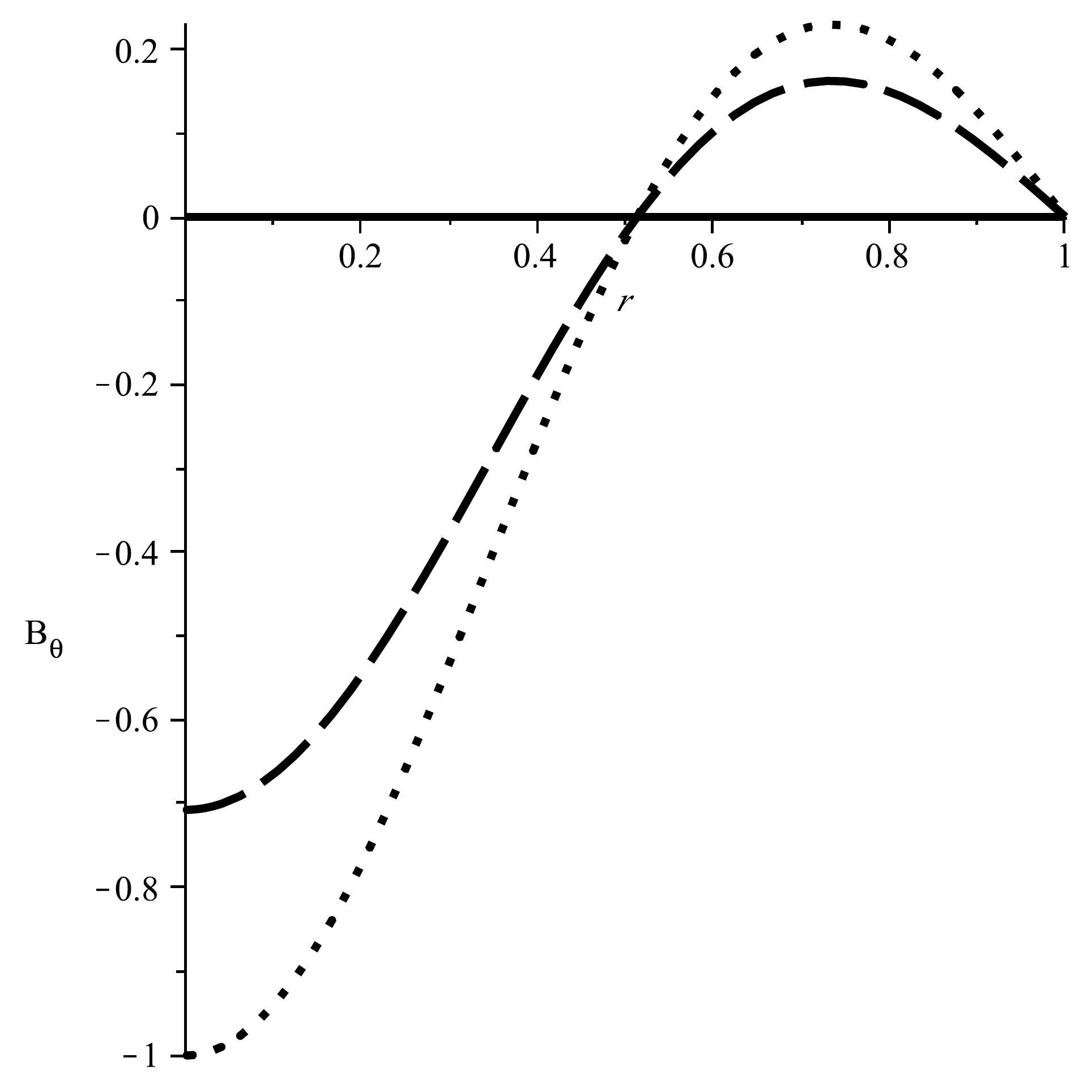}
\includegraphics[width=4.0cm]{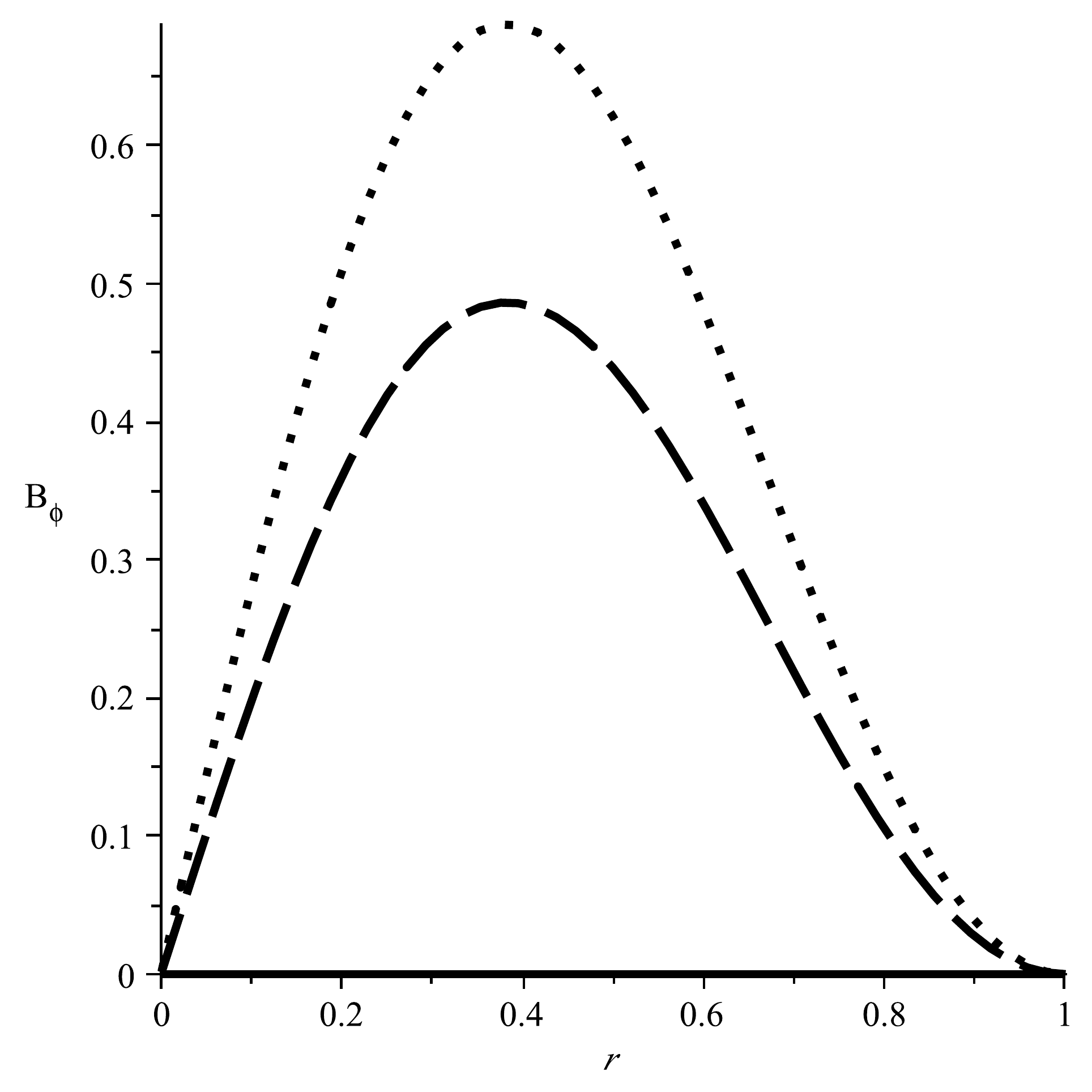}
\caption{Plot of the field components $B_{r}$, $B_{\theta}$ and $B_{\phi}$ along $r$ for $\theta=0$ (the axis), solid line, $\theta=\pi/4$, dashed line and $\theta=\pi/2$ (the equator). All three components go to zero at $r=1$ and in addition $B_{r}$ and $B_{\phi}$ reach $r=1$ with zero derivatives.}
\label{Bcomponents}
\end{figure}

\section{Physical implications}

\begin{figure}
\includegraphics[width=0.8\linewidth]{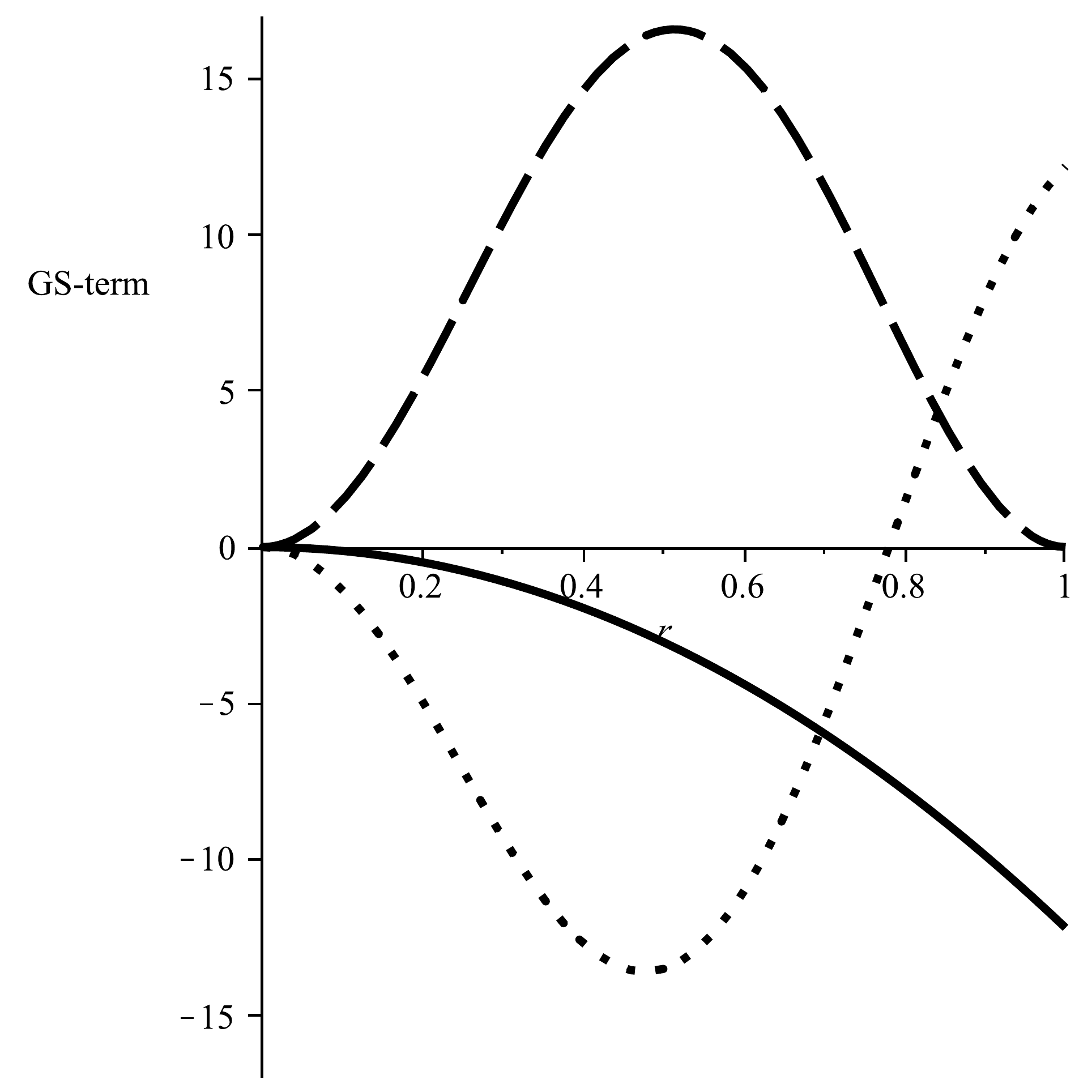}
\caption{Plot of the relative intensity of the terms that appear in the Grad-Shafranov equation~(\ref{GS}) in a section of the system at $\theta=\pi/4$. The solid line corresponds to term arising from the pressure, the dotted line to the poloidal tern and the dashed line to the toroidal term. The three terms add up to zero.}
\label{Force}
\end{figure}

We shall examine the relative contribution of the forces. The various terms in the Grad-Shafranov equation~(\ref{GS}) are proportional to the forces acting in the system. The term arising from the action of the differential operator $\Delta^{\ast}$ is due to the poloidal field force, the one containing $F(P)$ is due to the pressure gradient and finally the one containing $G(P)$ is due to the toroidal field force, Fig.~(\ref{Force}). 

\begin{figure}
\includegraphics[width=0.8\linewidth]{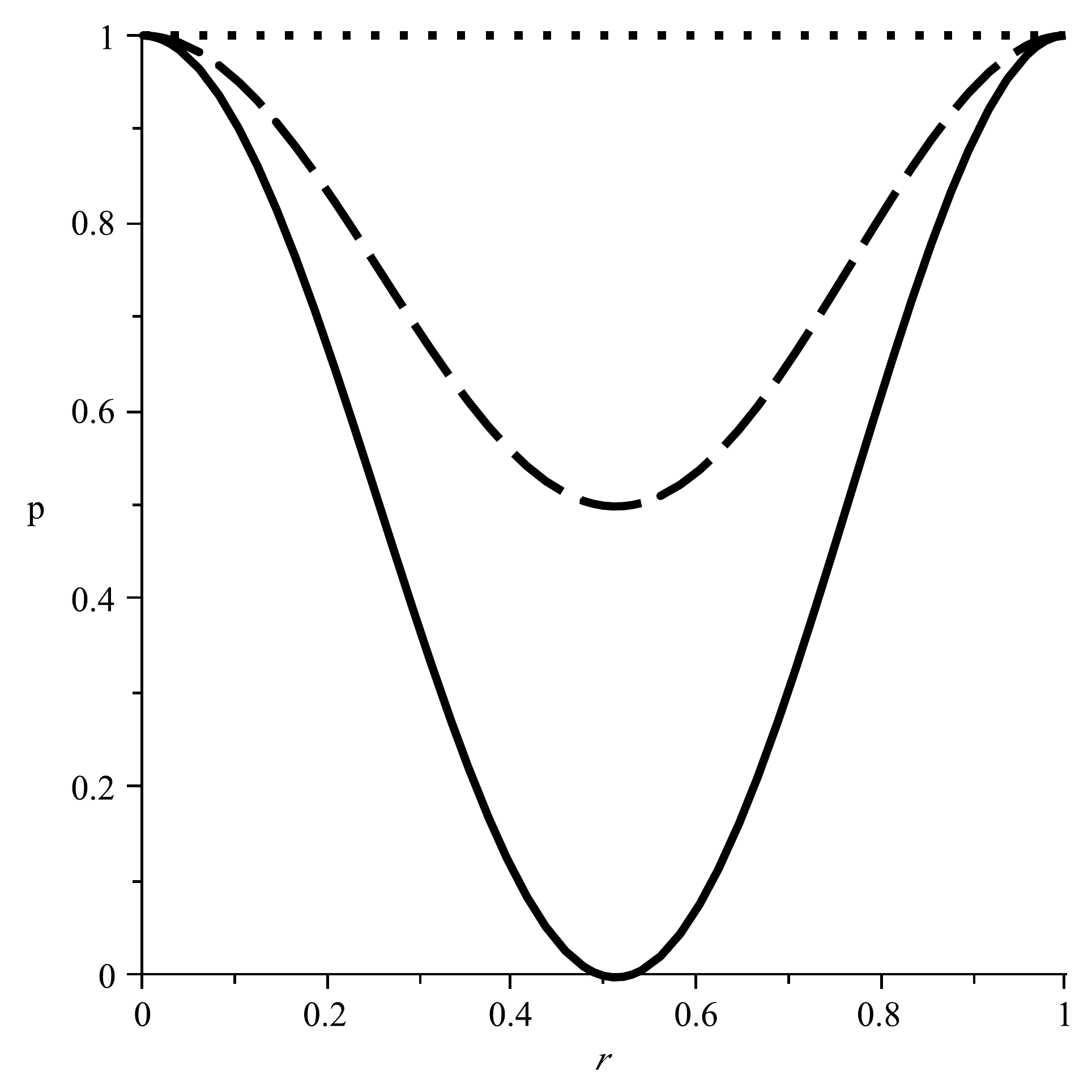}
\caption{The gas pressure $p$ as function of $r$; the solid line is for $\theta=\pi/2$, the dashed line is for $\theta=\pi/4$ and the dotted line is the pressure along the axis for $\theta=0$. We have chosen $p_{0}$ so that the minimum of the pressure is zero.}
\label{Pressure}
\end{figure}

The solution we have found determines only the pressure gradient and not its actual value. The pressure profile of the system is related to the solution through the relation $F=4\pi {\rm d} p/{\rm d} P$, given that we know $F_{0}$ the actual pressure is $p=p_{0}+F_{0}P/(4\pi)$, $p_{0}$ is a constant corresponding to the ambient pressure of the medium surrounding the cavity. The term $F_{0}P$ is always negative, irrespective of the normalization, thus the pressure always has a dip inside the cavity and it requires a positive term $p_{0}\geq (-F_{0}P/(4\pi))_{max}$, so that we never have a negative energy density. The gas pressure is plotted in Fig.~(\ref{Pressure}). 

We have evaluated the energy of the system as a function of $\alpha$, while keeping the radius fixed. Our aim is to find if the case without surface currents has any particular significance in terms of energy. The energy comprises three terms, one due to the magnetic field, a second one that is related to the term $F_{0}$ and a third term which is due to the ambient pressure $p_{0}$. Although the force equation and the solution do not contain the polytropic index ($\gamma$), this index appears in the energy calculation. We are evaluating the quantity $W$ which is the energy difference from the energy of a sphere of the same radius containing only gas of pressure $p_{0}$
\begin{eqnarray}
W=\frac{1}{8 \pi}\int B^{2}dV+\frac{1}{\gamma-1}\int (p-p_{0}) dV\,. 
\end{eqnarray}
A case of particular interest is for $\gamma=4/3$, as the magnetic field behaves like a fluid of $\gamma=4/3$. We have evaluated $W$ subject to given helicity, for a range of values of $\alpha$, but keeping the form of $F_{0}$ so that the system is confined within a sphere of given radius, see Fig.~5. For particular values of $\alpha$ the surface current vanishes.

We have found that when $\gamma=4/3$ the case without surface currents is the one that corresponds to minimum $W$ which is actually equal to zero, meaning that the energy contribution of the magnetic field and of the term containing $F_{0}$ are equal and opposite. This is a consequence of the fact that the magnetic field has an effective $\gamma=4/3$. Had we chosen an other polytropic index for the fluid the partition of energy in the fluid and the magnetic field would be different. 
\begin{figure}
\includegraphics[width=0.8\linewidth]{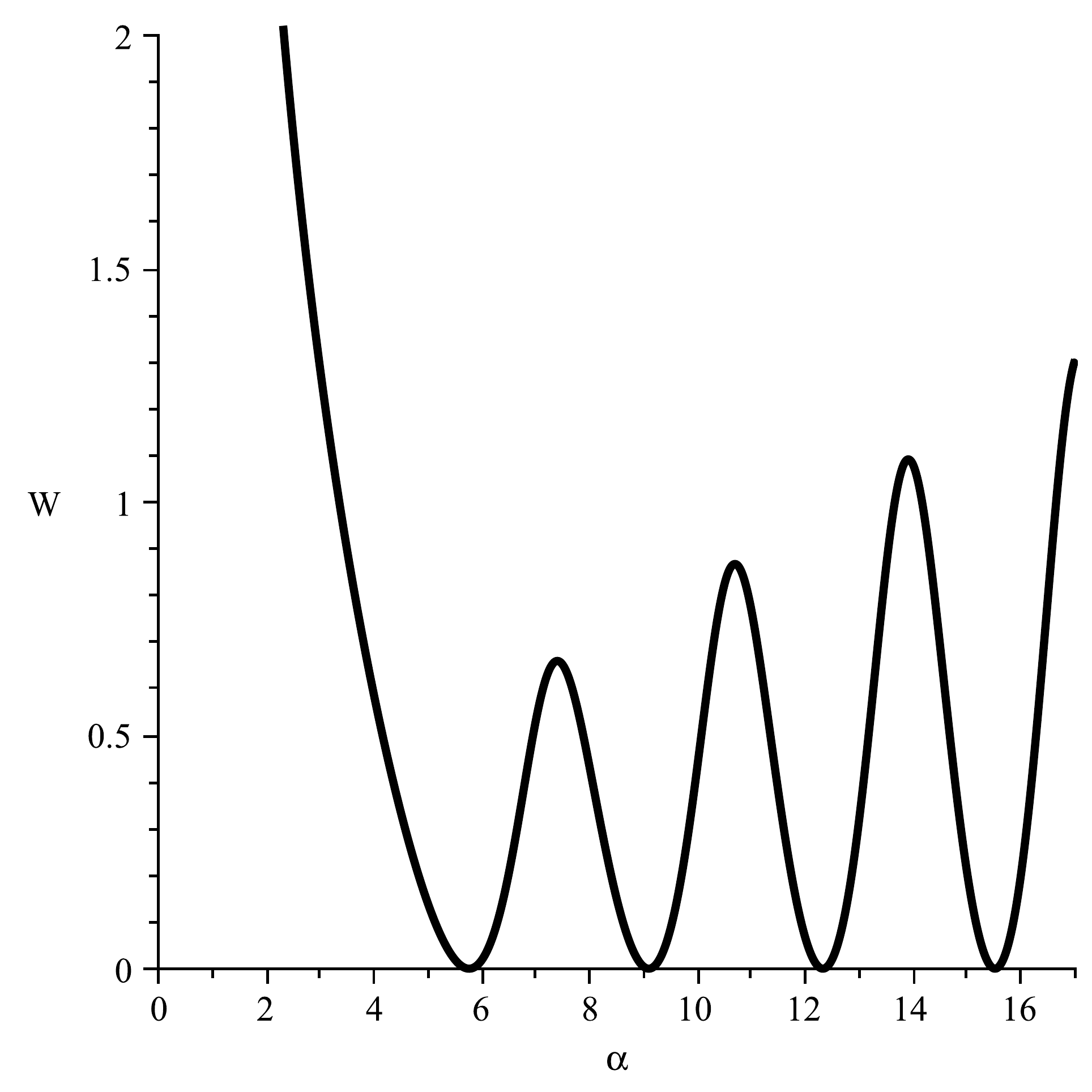}
\caption{$W$ as a function of $\alpha$ for a sphere of given radius, flux and helicity; and for a fluid of $\gamma=4/3$. The units are arbitrary. The minima occur at $\alpha$ which lead to systems without surface currents. The first minimum is at $\alpha=5.76$ which is the case we study in more detail.}
\label{W}
\end{figure}

In general, pressure balance requires that any equilibrium satisfies
\be
\frac{1}{24 \pi}\int B^{2}dV+\int (p-p_{0}) dV=0, \label{pressure-balance}
\ee
since, averaged over all directions, the magnetic field exerts a pressure equal to one third of its energy density -- just like any other relativistic fluid. This is confirmed by numerical experiments 

\section{Simulation}

Numerical simulation is the most straightforward method of testing the stability of this magnetic field configuration. In this section, such simulations are described.

The code used is the {\sc stagger code} \citep{NG1995, GN2005}, a high-order finite-difference Cartesian MHD code which uses a `hyper-diffusion' scheme, a system whereby diffusivities are scaled with the length scales present so that badly resolved structure near the Nyquist spatial frequency is damped whilst preserving well-resolved structure on longer length scales. This, and the high-order spatial interpolation and derivatives (sixth order) and time-stepping (third order) increase efficiency by giving a low effective diffusivity at modest resolution ($144^3$ here). The value of the diffusivity is set to somewhat more than the minimum required to avoid numerical instability in a variety of situations; since the scheme is designed primarily to damp behaviour at or near the Nyquist frequency, its effect on global structures is minimal -- the magnetic diffusivity present has a negligible effect on the magnetic energy during the simulations. In effect, the diffusive timescale in this static configuration is at
least three orders of magnitude greater than the Alfv\'en timescale. The code includes Ohmic as well as thermal and kinetic diffusion. The use of Cartesian coordinates prevents any problem of singularities and simplifies the boundary conditions: periodic boundaries are used here.

The numerical model is set up as follows. The computational box is a cube of side $3R$ where $R$ is the radius of the bubble. The boundaries are sufficiently far from the bubble that they have no significant effect. The bubble is embedded in an ambient pressure $p_0$ and can have arbitrary field strength up to some maximum where the plasma-$\beta$ is of order unity; for convenience we choose a field strength such that the minimum gas pressure inside the bubble is half that outside, i.e.  $p_{\rm min}= p_0/2$. So that all potentially unstable modes are excited, a random perturbation of strength $1$\% containing a range of all relevant length scales (with a flat spectrum) is given to the pressure field. The fluid is then evolved forwards in time: stability can be assumed if nothing happens during a number of Alfv\'en timescales, since any instability should grow on this timescale.

\begin{figure}
\includegraphics[width=1.0\hsize]{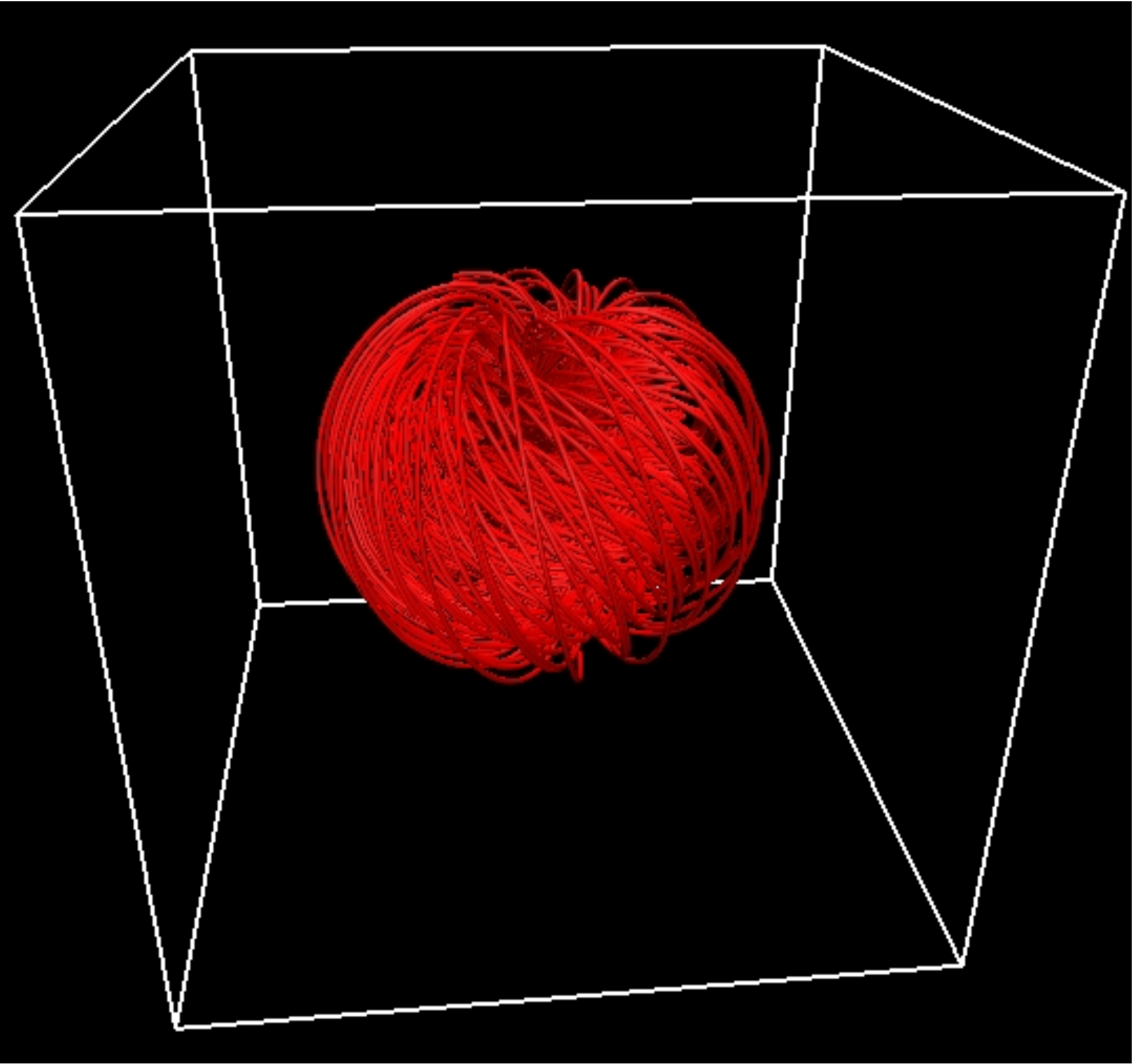}
\caption{Three-dimensional rendering of the equilibrium field embedded in the computational box.}
\label{vapor}
\end{figure}

\begin{figure}
\includegraphics[width=1.0\hsize]{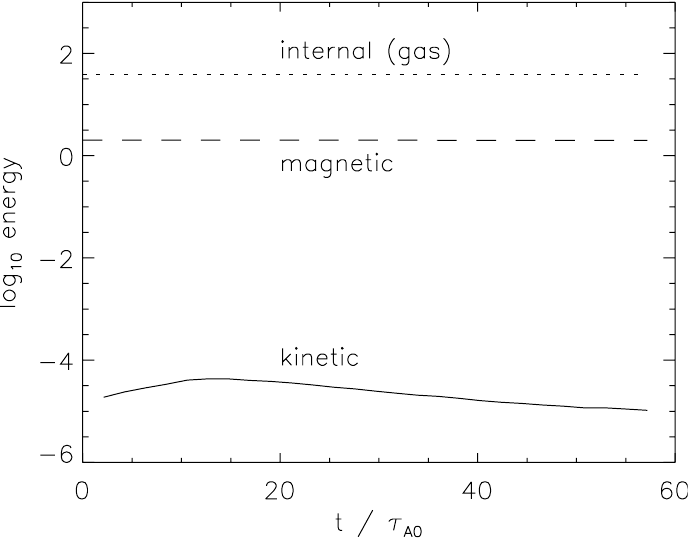}
\caption{Kinetic, thermal and magnetic energies as a function of time. Time is expressed in units of the Alfv\'en crossing time. The initial growth in kinetic energy comes from the perturbation to the pressure field; the lack of any further growth implies stability.}
\label{energies}
\end{figure}

The computational setup and magnetic field are illustrated in Fig.~\ref{vapor}. There is no appreciable change in the magnetic field during the simulation. In Fig.~\ref{energies} are plotted the time evolution of the kinetic, thermal and magnetic energies. Clearly, the kinetic energy does not grow and none of the excited modes is unstable. Since all imaginable modes are excited initially, the only logical conclusion is that the configuration described in the previous section is indeed stable.


As mentioned above, there is a limit on the strength of this field; if the field strength is increased at constant ambient pressure $p_0$ then eventually zero gas pressure is reached on the neutral line. This happens when the mean plasma-$\beta$ in the bubble is approximately unity. Configurations with low-$\beta$ must have a different form: since the interior of any low-$\beta$ bubble must be roughly force-free, the magnetic field is confined by a boundary layer with a strong pressure gradient. We can investigate this type of equilibrium by performing a simulation similar to that above but with reduced gas pressure in the bubble interior. The simplest setup is to use for the initial conditions a magnetic field of the same form as before but with increased strength and a low, uniform gas pressure inside the bubble, whilst satisfying the pressure balance condition (\ref{pressure-balance}), and to follow the relaxation of the configuration into a low-$\beta$ equilibrium. In the following simulation, the bubble has a gas pressure of $0.1p_0$ so that the isotropic magnetic pressure $B^2/24\pi=0.9p_0$. Of course, this is a rather crude way of producing a low-beta equilibrium and a more realistic method would involve a self-consistent cooling mechanism; however the details do not affect the qualitative conclusions.

\begin{figure}
\includegraphics[width=1.0\hsize]{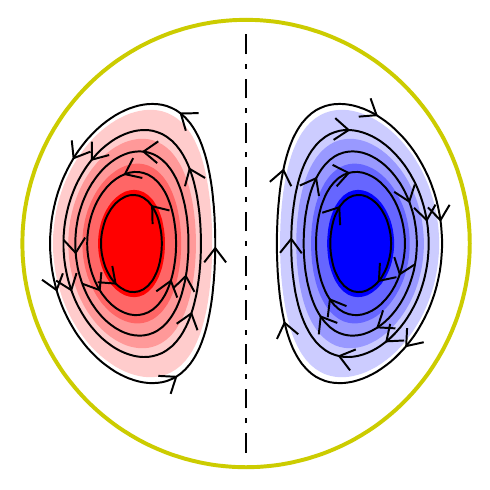}
\includegraphics[width=1.0\hsize]{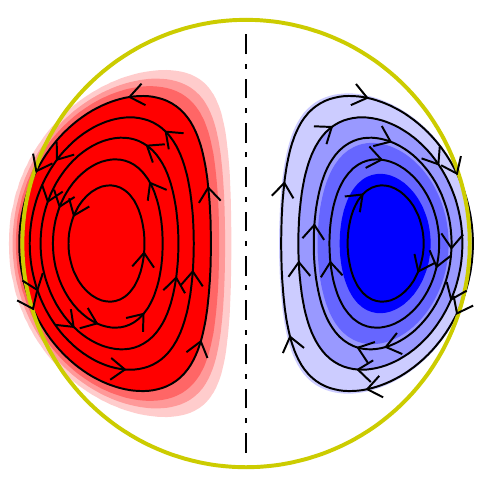}
\caption{{\it Upper panel}: cross-section of the high-$\beta$ equilibrium. Black lines represent poloidal field; on the right-hand-side the blue shading represents the toroidal field (multiplied by the cylindrical radius $r\sin\theta$) and on the left-hand-side the red shading represents gas pressure. The thick yellow line represents the boundary of the bubble. That both $B_\phi r\sin\theta$ and $p-p_0$ are functions of the flux function and are therefore constant on poloidal field lines is clear. {\it Lower panel}: the low-$\beta$ equilibrium. The pressure is very low throughout the bulk of the bubble and there is a steep pressure gradient near the boundary which balances the strong Lorentz force in that region; note that the field strength now no longer goes to zero so gently at the boundary. Note also that during relaxation the shape of the bubble has changed somewhat to an oblate spheroid, although the volume has remained constant.}
\label{xsec}
\end{figure}

\begin{figure}
\includegraphics[width=1.0\hsize]{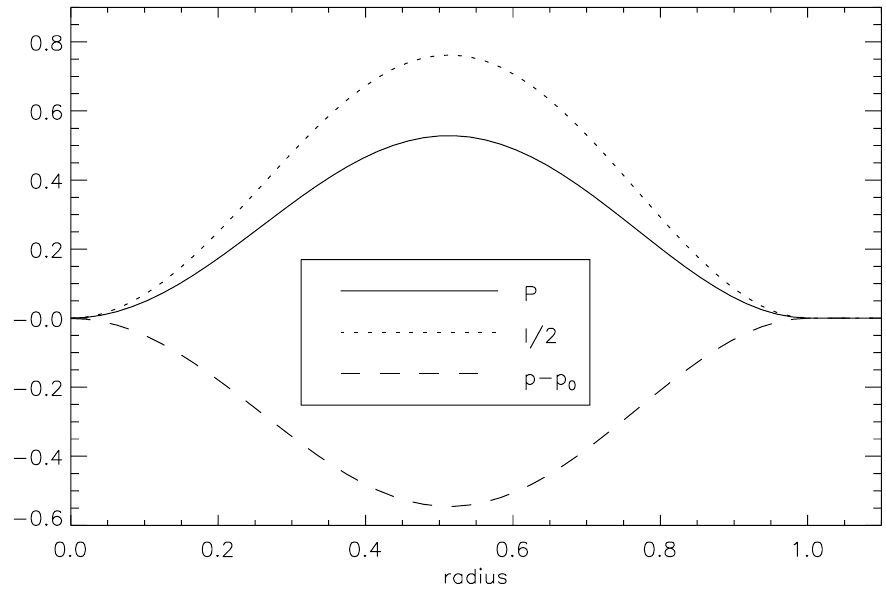}
\includegraphics[width=1.0\hsize]{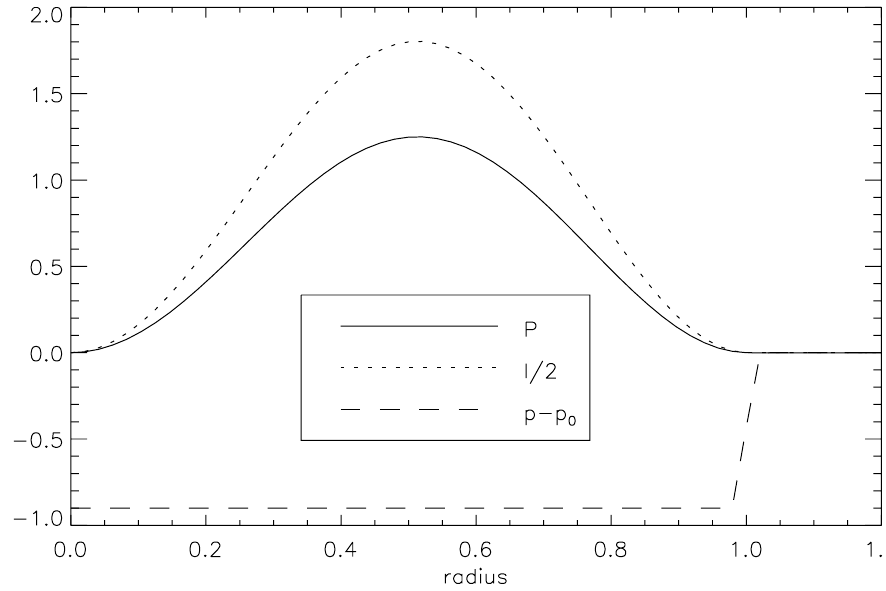}
\includegraphics[width=1.0\hsize]{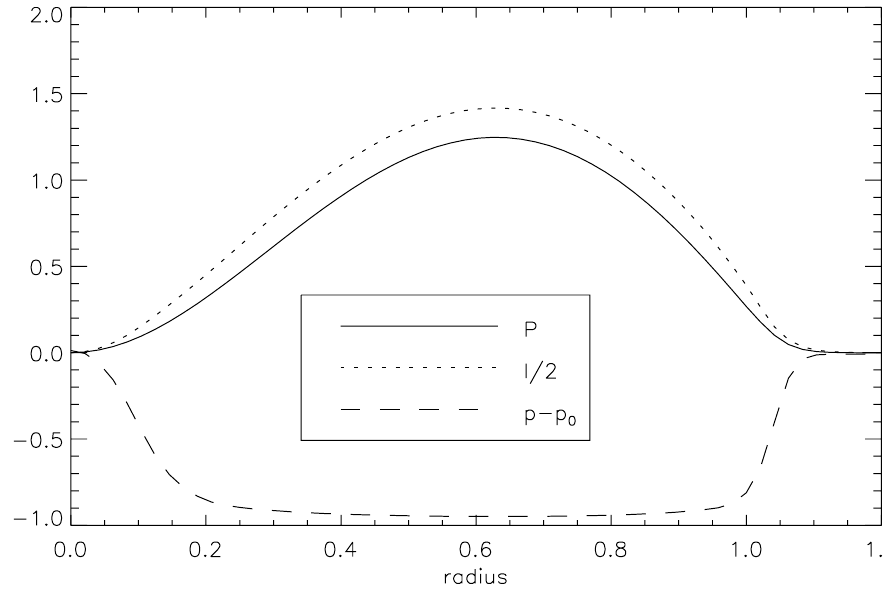}
\caption{{\it Upper panel}: values of the flux function $P$, toroidal field $\times$ cylindrical radius $B_\phi r\sin\theta$ and pressure difference $p-p_0$ in the equatorial plane of the high-$\beta$ equilibrium. {\it Middle panel}: the initial conditions of the low-$\beta$ simulation. 
 {\it Lower panel}: the new low-$\beta$ equilibrium after relaxation.}
\label{cut}
\end{figure}

When the simulation is set running, dynamical motions appear and an equilibrium is approached. To aid relaxation into the new equilibrium, an artificial friction force of appropriate magnitude is added to the momentum equation so that oscillations about the new equilibrium are damped. In Fig.~\ref{xsec} the structures of the original high-$\beta$ and the new low-$\beta$ equilibria are compared. The shape of the bubble has deviated from spherical\footnote{There is probably a general tendency for axisymmetric equilibria to be oblate spheroidal, as this shape allows the poloidal lines to be more nearly circular, circular being a lower energy state. A proper investigation of non-spherical equilibria is left for the future.}, and the bulk of the interior is now approximately force-free. A boundary layer of high pressure gradient confines the magnetic pressure, and since the gas pressure must be constant along field lines, there is a narrow region of high gas pressure around the axis of symmetry. Various quantities are plotted in Fig.~\ref{cut} on the equatorial plane.

Physically it seems unlikely that a low-$\beta$ equilibrium could form directly from relaxation of the AGN ejecta or whatever inflated the bubble. This is because during the relaxation from an arbitrary or `turbulent' injected magnetic field, most of the magnetic energy is transferred to the gas, going into some combination of thermal energy and relativistic particles; in any case, a high-$\beta$ equilibrium is the result. A low-$\beta$ equilibrium could however form later if the bubble gas is able to cool radiatively, in fact not unlike the formation in this simulation except that a gradual loss of thermal energy would cause quasi-static evolution of the equilibrium rather than the sudden change effected here.

\section{Observational properties}

In this section we evaluate some observable properties of the system. ICM cavites are observed as depressions in the X-ray emission and also are associated with radio lobes, in addition other simulated works \citep{D2009,RM2005} have similar predictions. For that reason we have evaluated the rotation measure of the Faraday rotation the cavity will cause to some radiation that lies behind it, the synchrotron radiation and its polarization created by the cavity and the depression it causes to the X-ray emission. 

There are two ways to test the magnetic field structure in the cavity. In the case of a cavity filled with non-relativistic material, the structure will cause Faraday rotation to the radiation that comes from behind the cavity. This case is investigated in section \ref{RMS}. If the cavity is filled with relativistic electrons they will produce synchrotron radiation which shall be polarized, in addition relativistic particles do not contribute to the rotation measure. Finally we present synthetic X-ray images of the system. 

The solution we have found, contains information about the magnetic field and consequently about the pressure, however in order to evaluate the observable quantities we need to know the density and the temperature. It is indeed true that the properties of the cavities can be explained by assuming a much higher temperature and a low density inside the cavity. However this will allow a great freedom in the possible combinations which could lead to similar profiles. In our work, we have decided to assume that the density and the temperature are associated to the pressure both inside and outside the cavity by the same law. Therefore, in the case of the rotation measure we shall assume that the non-relativistic plasma is described by $\gamma=5/3$, whereas in the case of the synchrotron radiation and the synthetic X-ray images, the system is described has $\gamma=4/3$ and temperature depends on the pressure. 

 An other issue related to the observability of the quantities discussed above, is due to synchrotron emission and rotation measure from the ICM itself. The intracluster medium is denser and cooler compared to the cavity, however according to the magnetic description of cavities we are proposing in this paper, the magnetic field inside the cavity shall be stronger than the magnetic field of the intracluser medium, which in our idealised solution is negligible. In addition if we integrate the emissivity along the line of sight the typical cavity size shall contribute $\sim 10$kpc whereas the length of the intracluster medium the line of sight crosses, is an order of magnitude larger or more. Thus we have competing factors, as the stronger magnetic field shall enhance the cavity contribution, whereas the lower density and shorter length shall depress it.

\subsection{Rotation measure}
\label{RMS}

We have evaluated the rotation measure corresponding to the solutions we have found. The rotation measure is the integral of the magnetic field parallel to the line of sight multiplied by the electron density along the line of sight. 
\begin{eqnarray}
RM=\frac{e^{3}}{2 \pi m_{e}^{2} c^{4}}\int_{0}^{d}n_{e}B_{\|}dl\,.
\end{eqnarray}
Then we use this information and we draw a map. In this problem the magnetic field is described accurately by the solution. However, in order to estimate the density we need to make an assumption for a relation between the pressure and the density, in this paradigm we have chosen the density to obey an adiabatic relation of $\gamma=5/3$. We calculate the contribution to the rotation measure only from the cavity and not from the ambient medium. In reality the rotation measure may be dominated by the contribution of the ambient medium which is more extended and denser. Our prediction shall be superimposed to the overall profile of the ICM.

In addition, as it was described above the pressure is determined up to an additive constant, we have chosen this additive constant so that the pressure at its minimum inside the cavity to be zero, i.e. the $\beta\approx1$ case we have also tested cases where the dip in pressure inside the cavity is a fraction of the external pressure and we have found that the rotation measure is very similar to that of a spheromak configuration embedded in a uniform density and pressure environment.

\begin{figure}
\centering
\includegraphics[width=2.7cm]{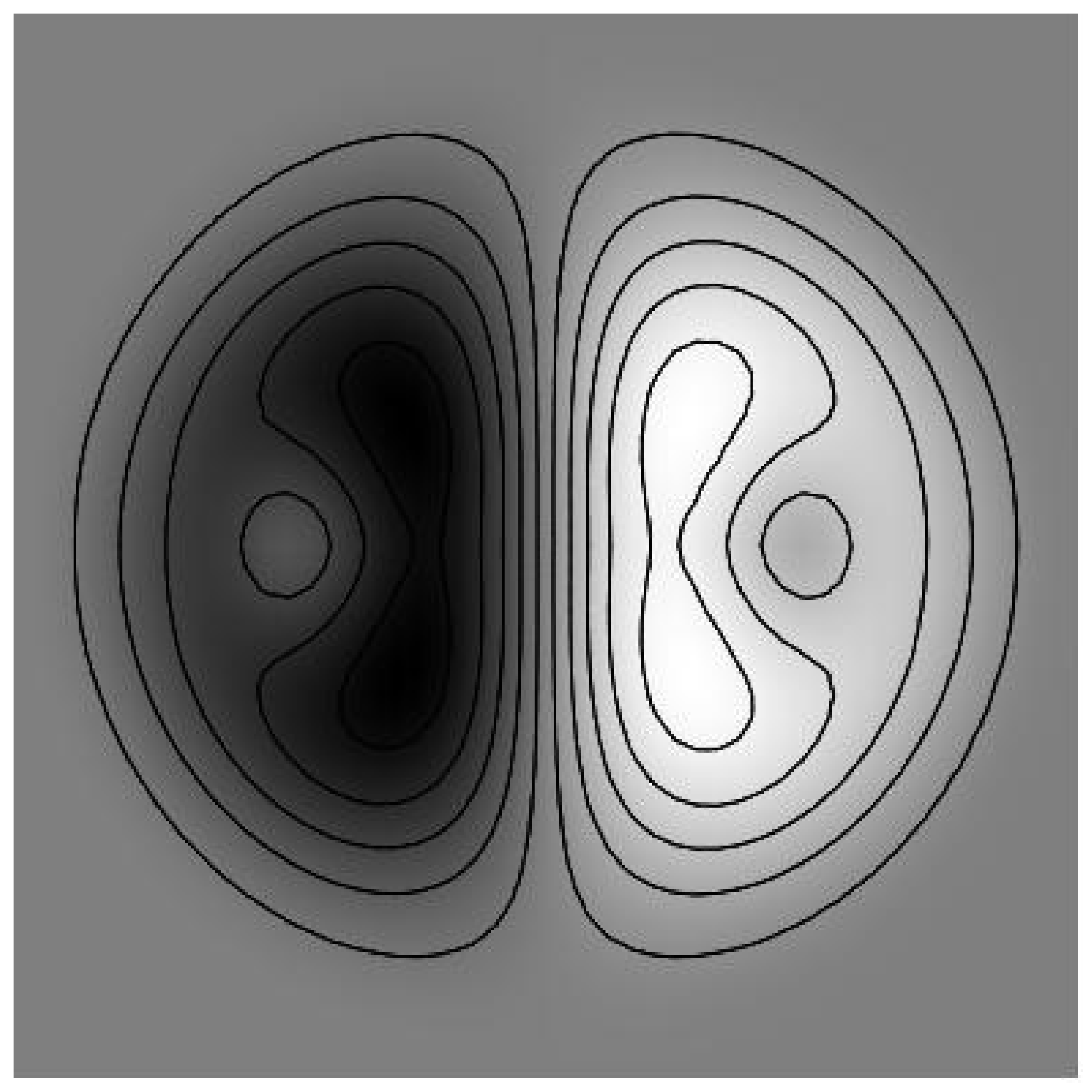}
\includegraphics[width=2.7cm]{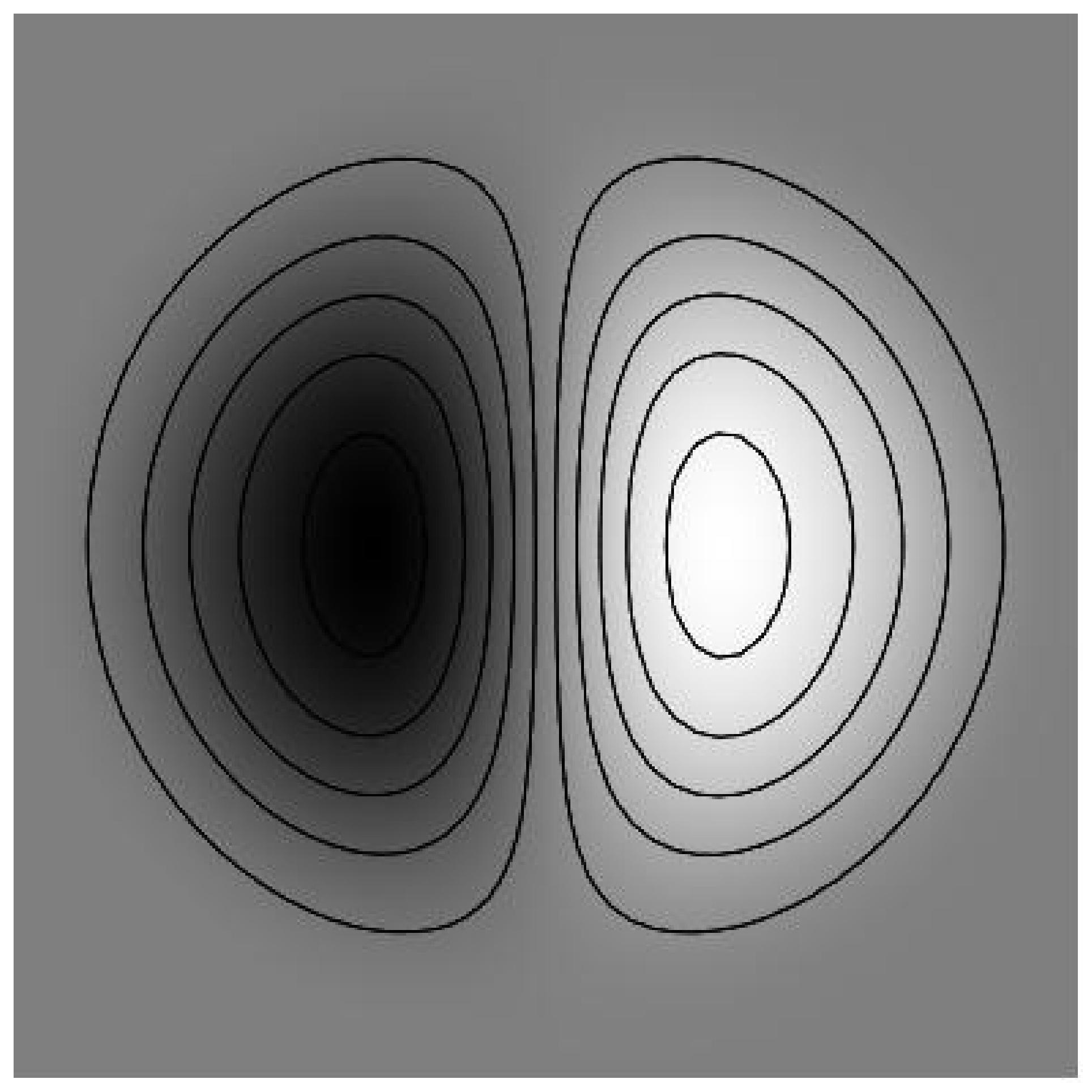}
\includegraphics[width=2.7cm]{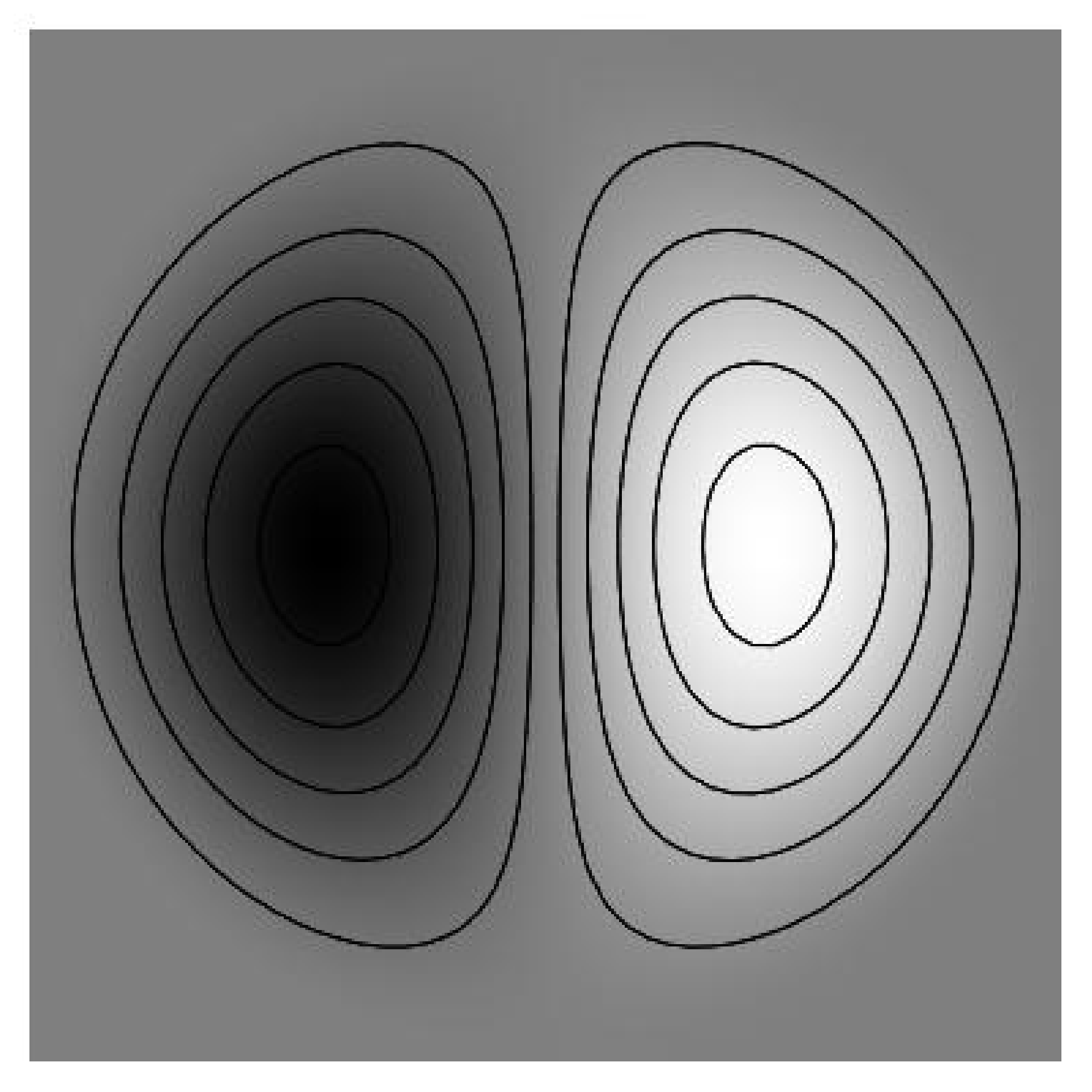}
\includegraphics[width=2.7cm]{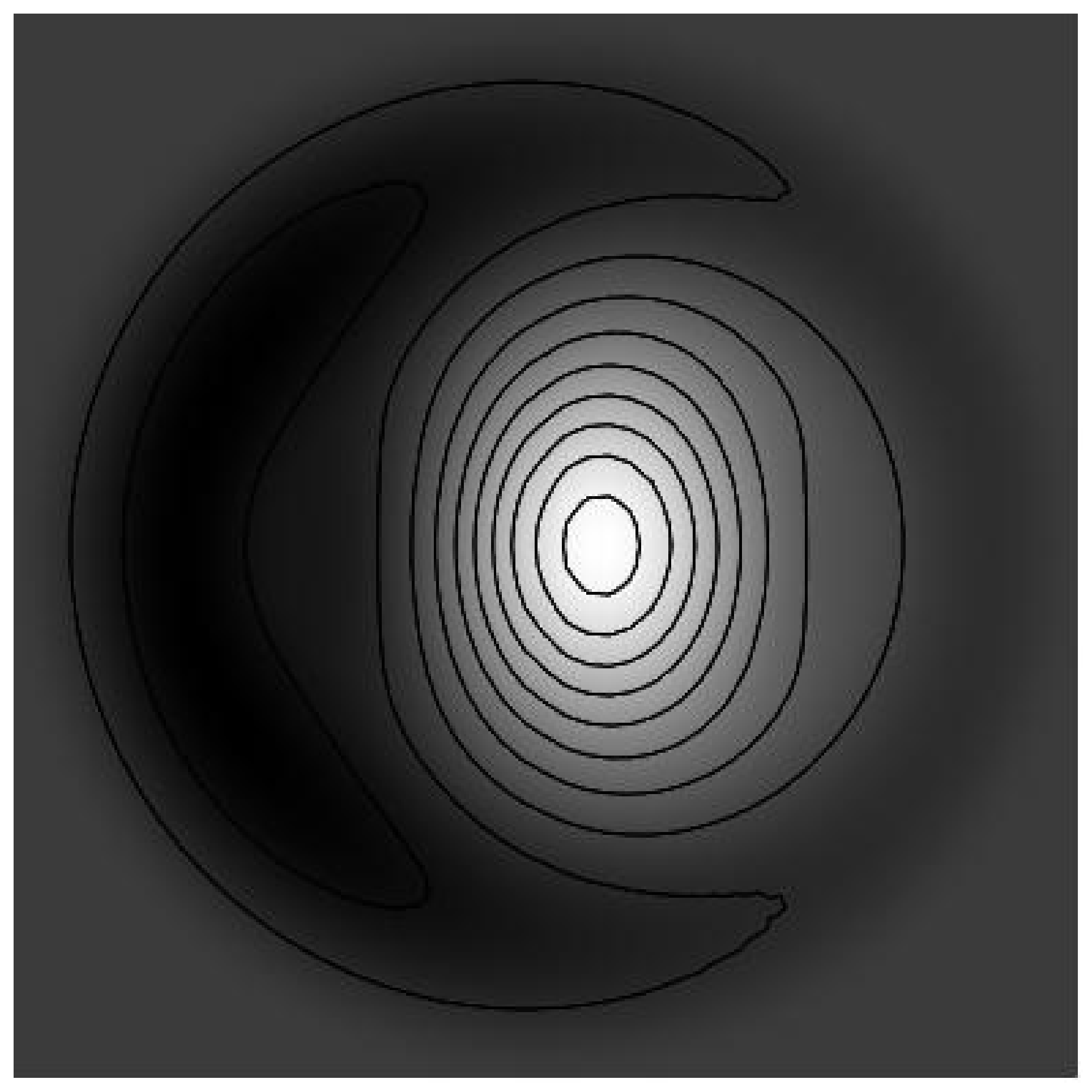}
\includegraphics[width=2.7cm]{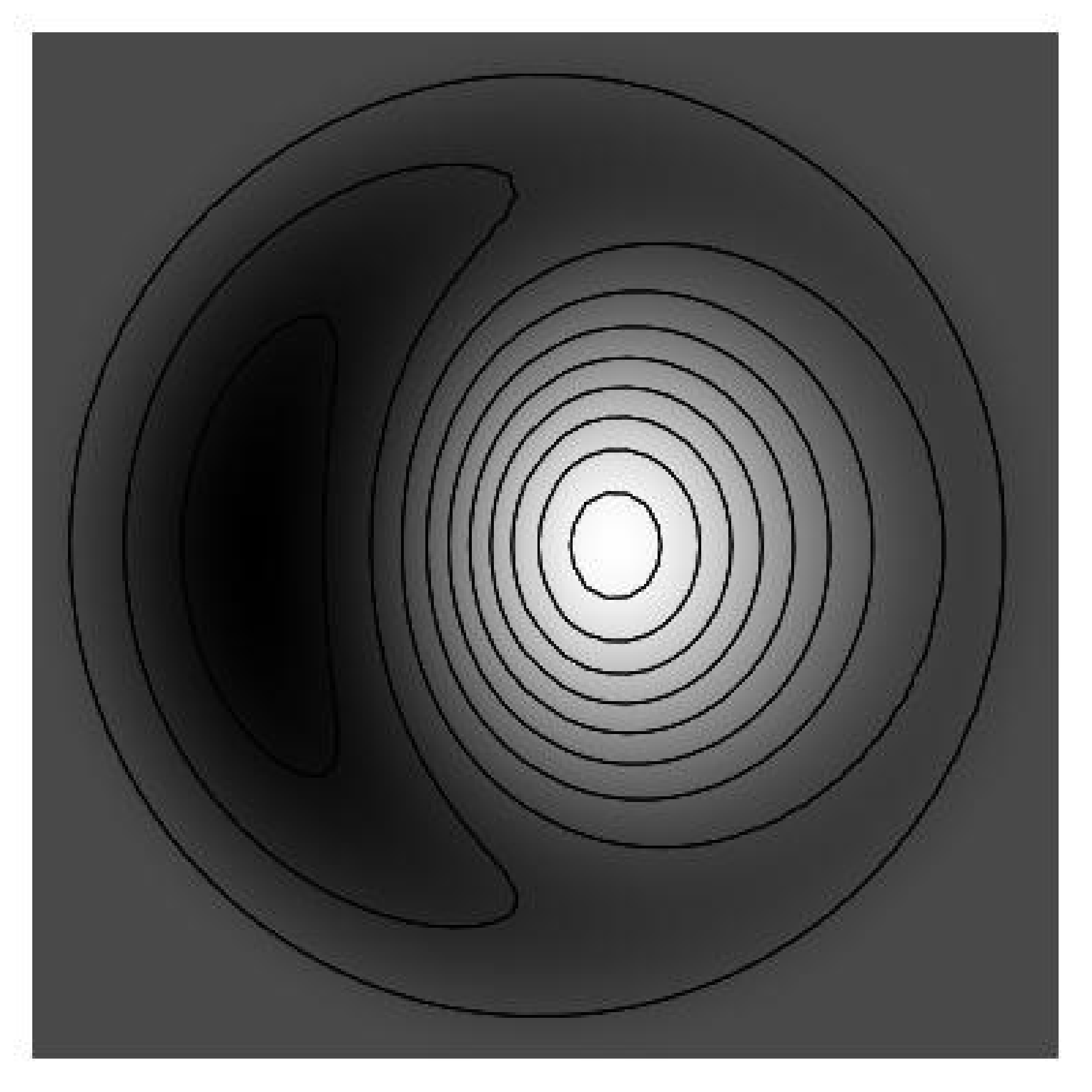}
\includegraphics[width=2.7cm]{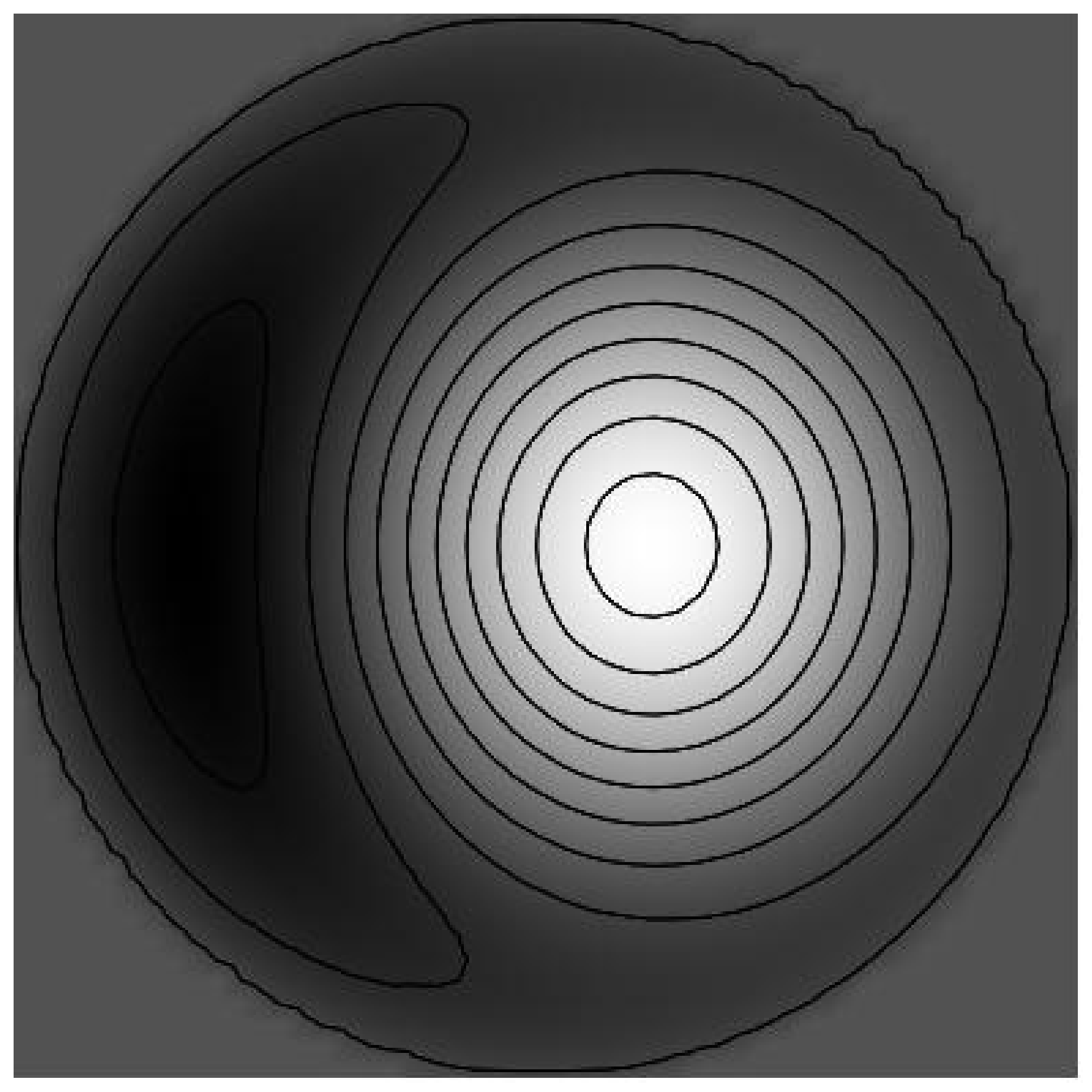}
\caption{The rotation measure for three models of cavities for two viewing angles, the first row is for the case where the line of sight is perpendicular to the axis of the system, whereas the second is for systems where the line of sight forms an angle of $\pi/4$ with the axis. The first column corresponds to the case where the minimum of pressure in the system is equal to zero. The second column is the system corresponds to the case  where the minimum of the pressure of the system at the minimum is half of the ambient pressure. The third column is a spheromak filled with gas of constant pressure.}
\label{RM}
\end{figure}

\begin{figure}
\includegraphics[width=.8\linewidth]{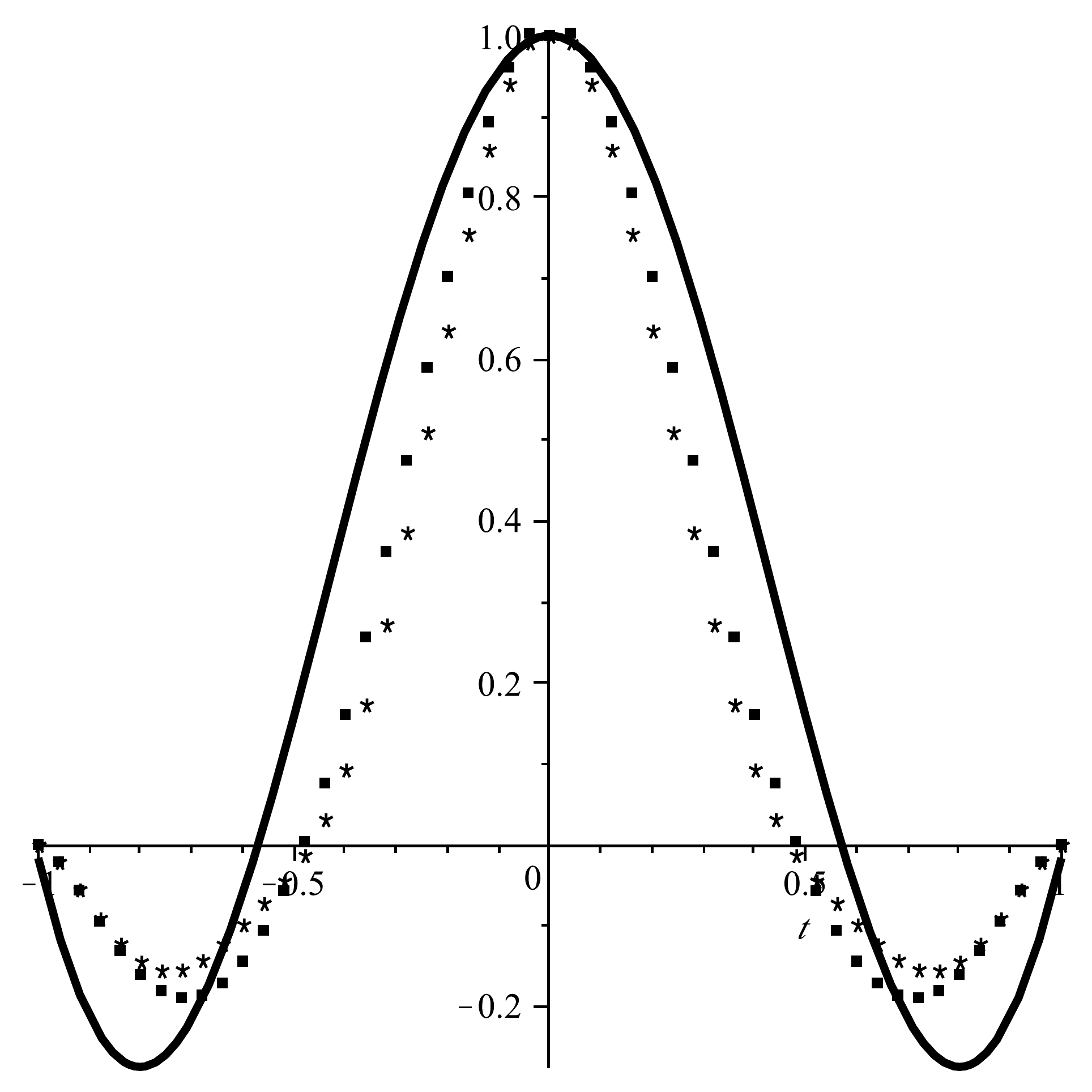}
\caption{The rotation measure for the case where the line of sight is along the axis of the system. In this case the rotation measure is radially symmetric, thus we plot a slice. The squares correspond to a solution with ambient pressure such as to go to zero at the minimum inside the cavity. The asterisks correspond to a solution with an ambient pressure twice as much as the previous case. Finally the solid line is the rotation measure for a spheromak embedded in a constant pressure environment.}
\label{RMA}
\end{figure}

We have studied three viewing angles with respect to the axis of the magnetic field. In the first case we have an equatorial view, the axis is perpendicular to the line of sight; in the second one the axis is tilted, the angle between the line of sight and the cavity axis is $\pi/4$; and in the third case, the polar view, the axis points towards the observer, Figs.~(\ref{RM}) and (\ref{RMA}). 

Coming to the actual value of the rotation measure it depends on the density of the electrons, the size of the cavity and the magnetic field. Setting the maximum value for density inside the cavity equal to $10^{-4}cm^{-3}$, the magnetic field $10\mu G$ and assuming a diameter of $10$kpc we find that the maximum absolute contribution of the rotation measure is about $10^{-3}rad/cm^{2}$. Thus if there are any electrons available to cause faraday rotation observations of sufficient resolution shall see a change in the rotation measure by $2\times 10^{-3}rad/cm^{2}$. Observations of the Perseus cluster \citep{dBB2005} suggest that these rotation measure values are within the limits of observation.


\subsection{Synchrotron emission and polarization}
\label{POLS}

\begin{figure}
\centering
\includegraphics[width=2.7cm]{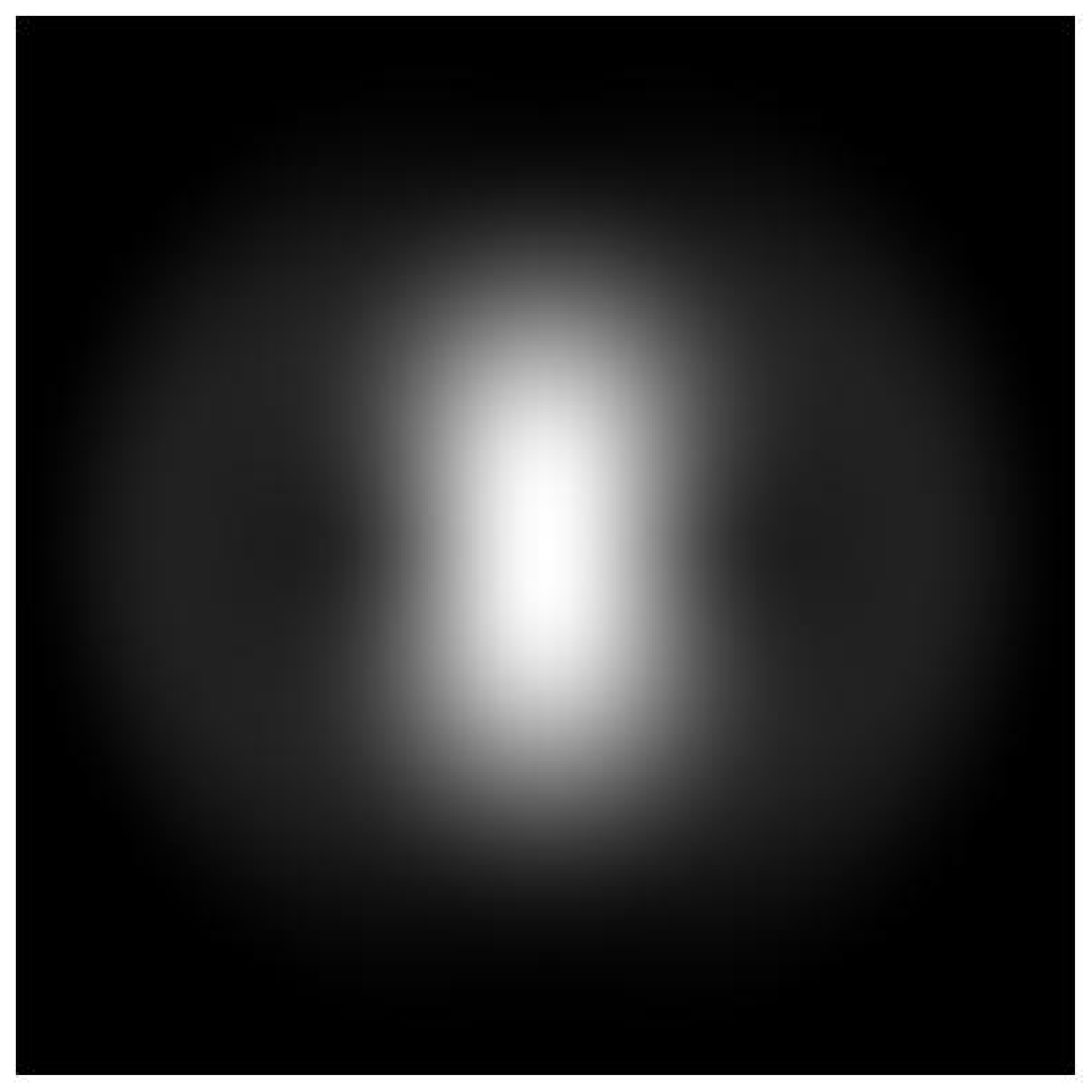}
\includegraphics[width=2.7cm]{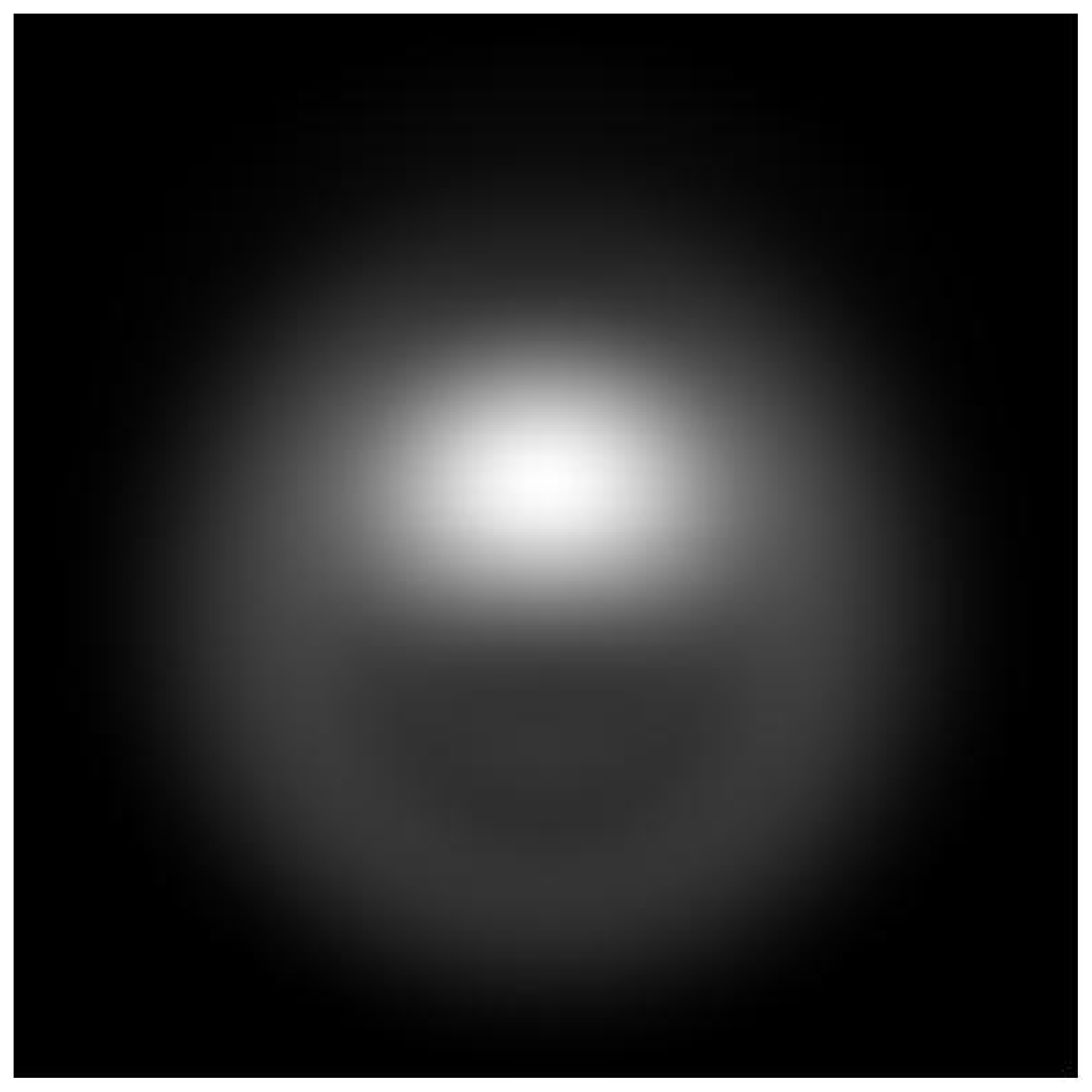}
\includegraphics[width=2.7cm]{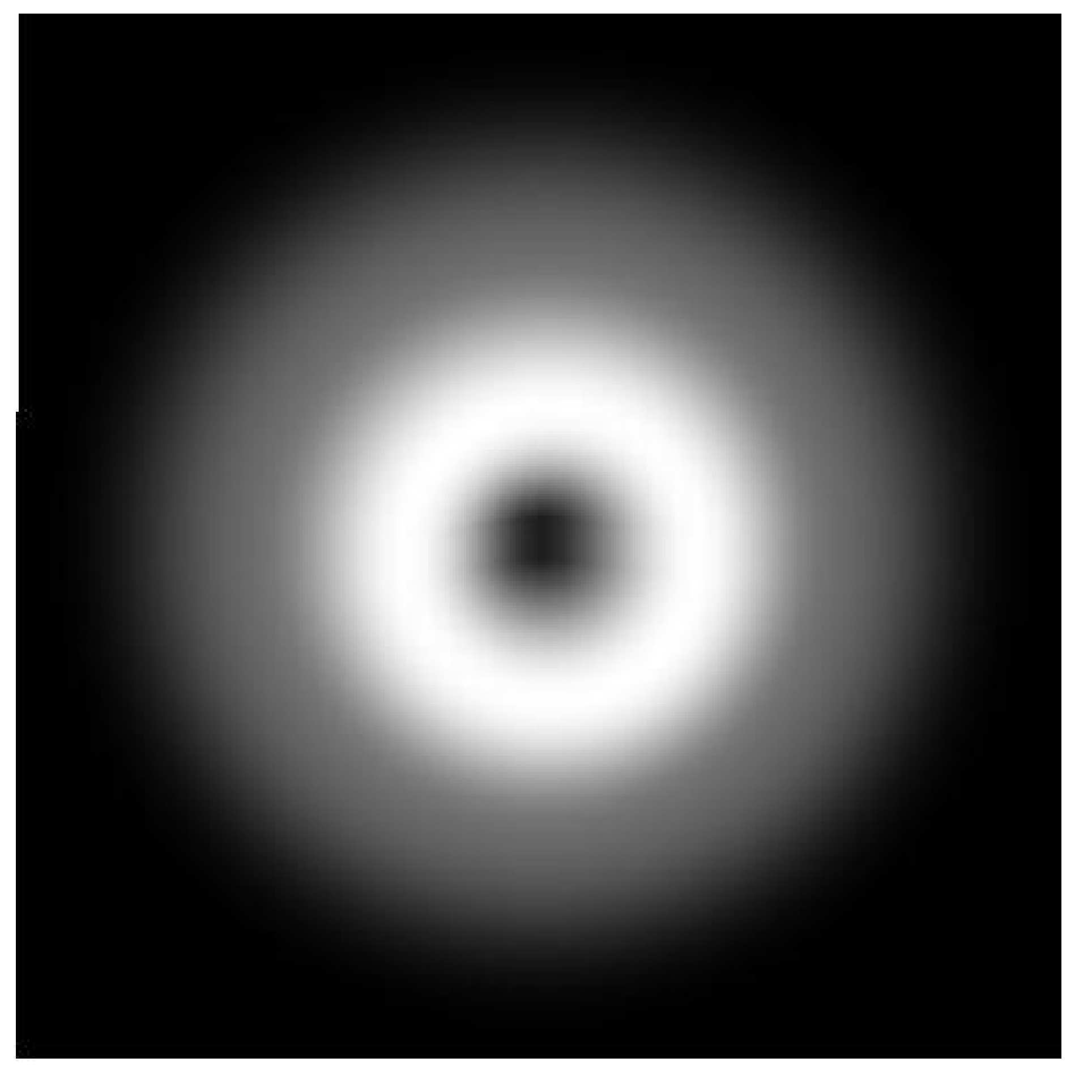}
\includegraphics[width=2.7cm]{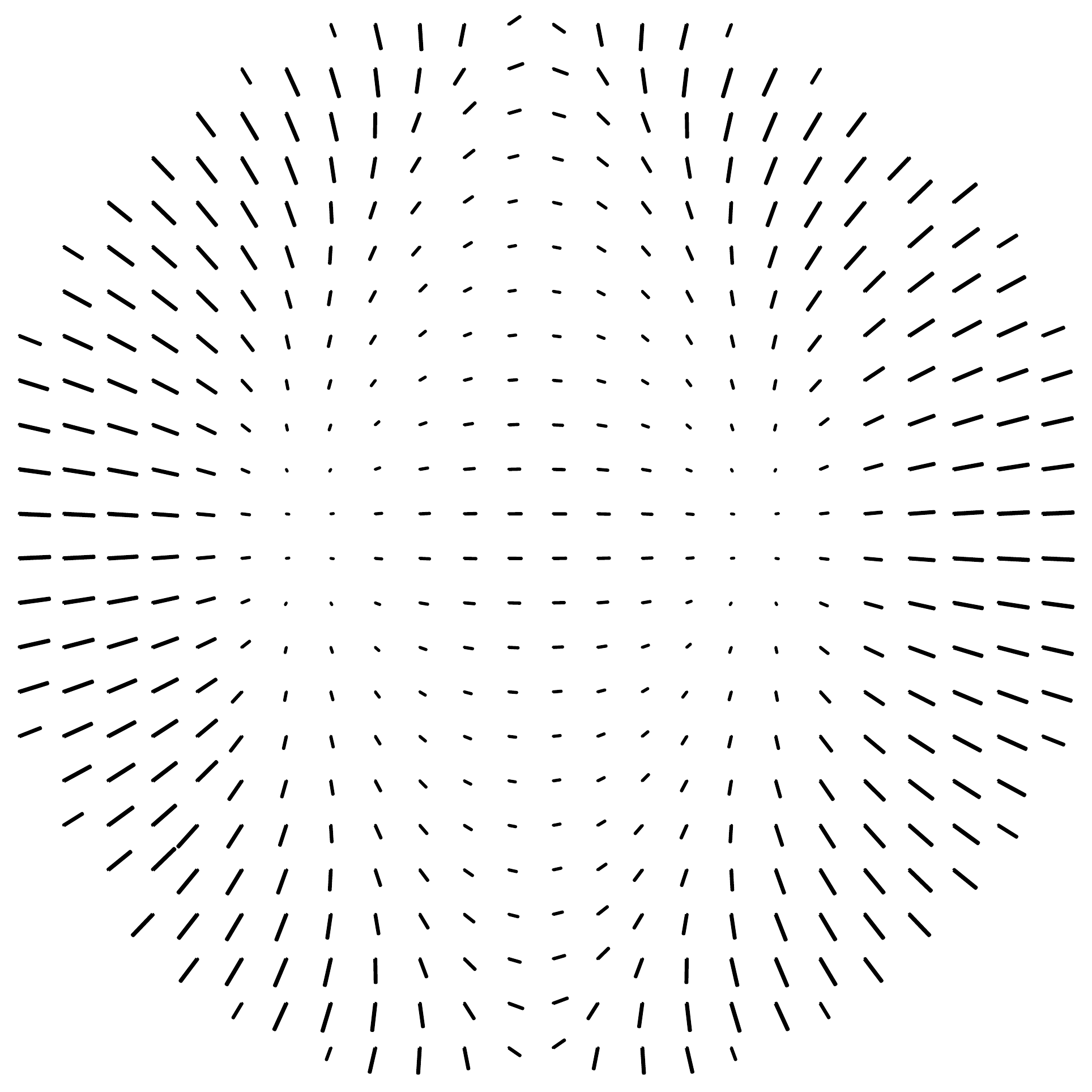}
\includegraphics[width=2.7cm]{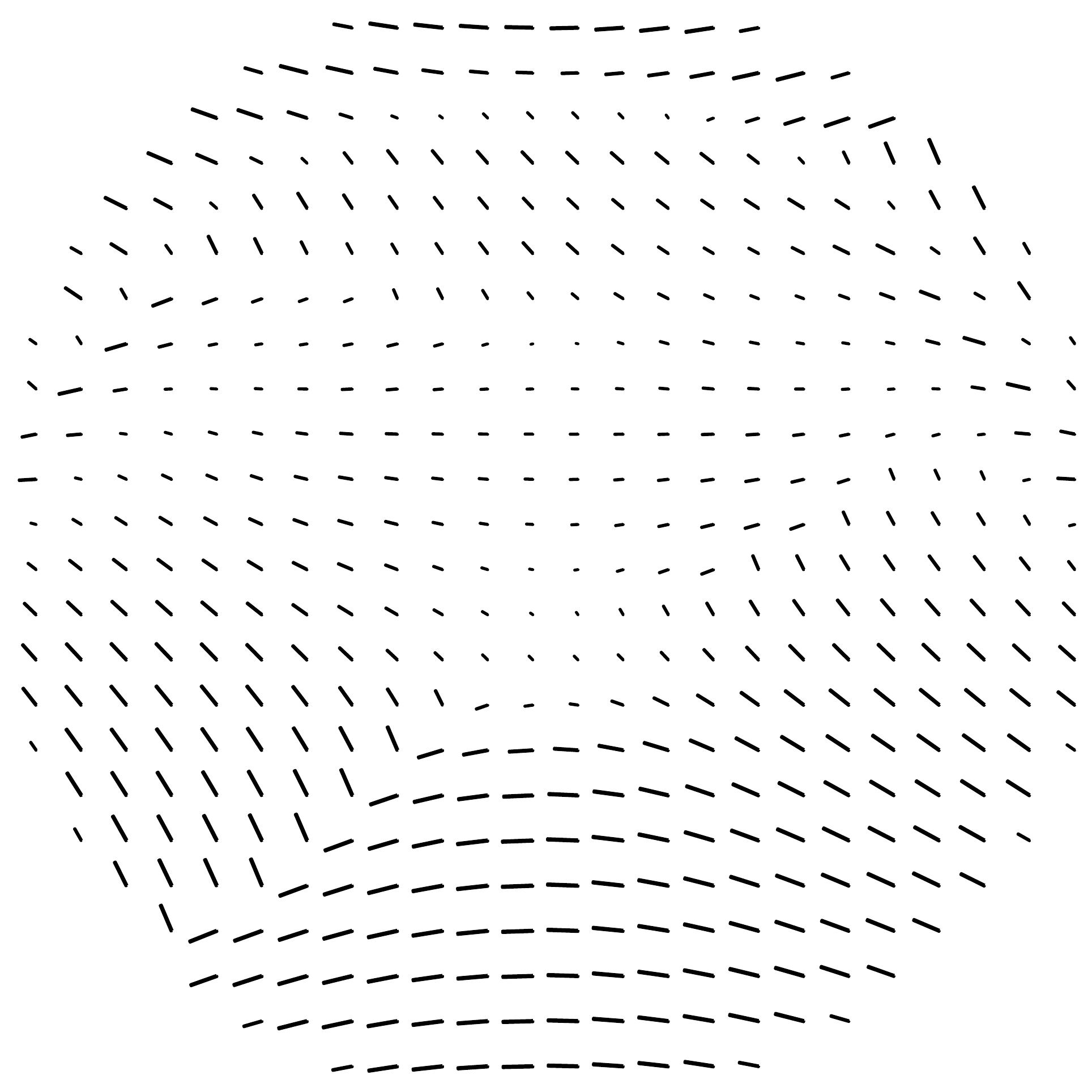}
\includegraphics[width=2.7cm]{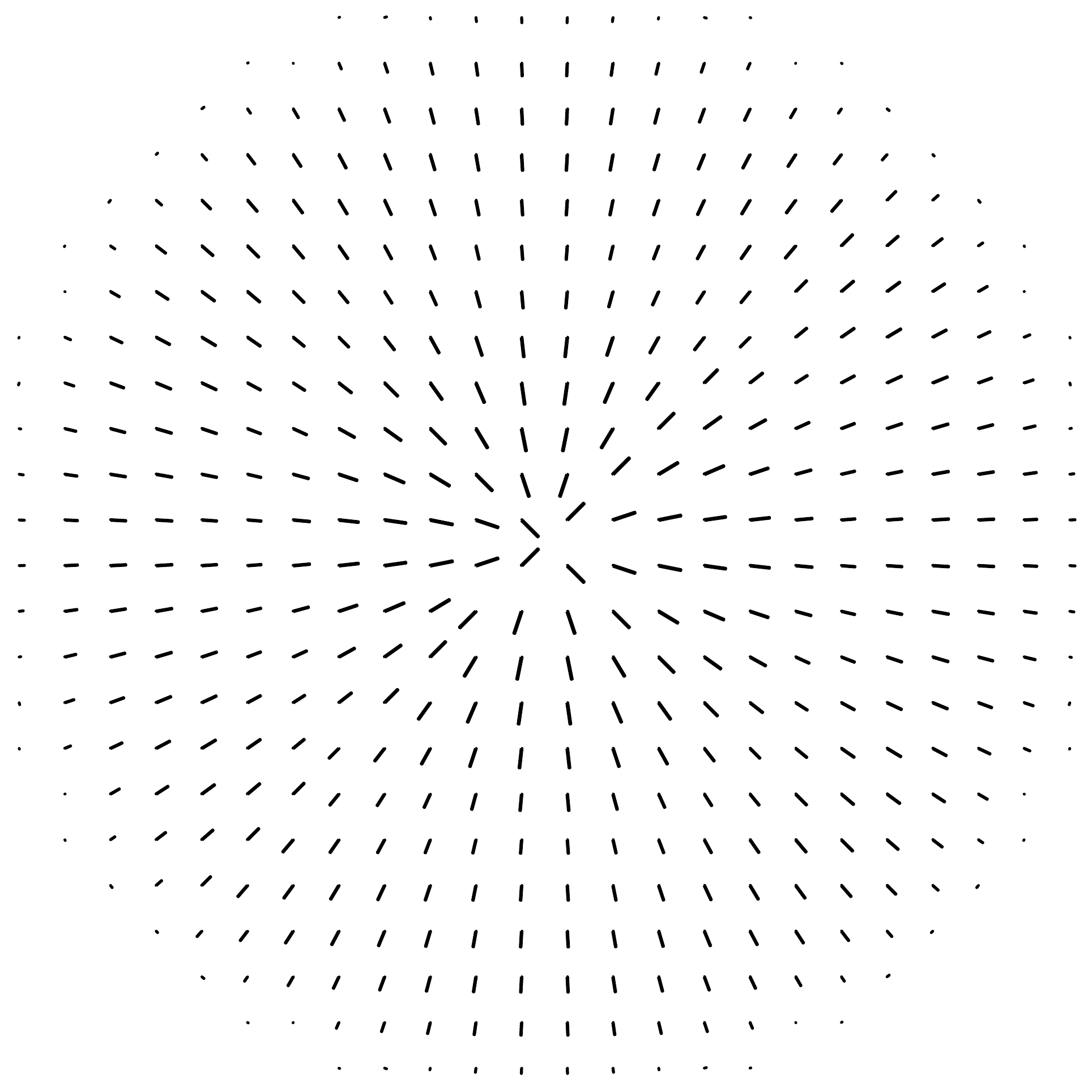}
\caption{First Row: The intensity of the synchrotron emission of the system for a system where the line of sight and the axis of the cavity form an angle of $\pi/2$, $\pi/4$ and $0$ from left to right respectively. Brighter areas have greater emissivity. Second Row: The polarization of the synchrotron emission again for angles of $\pi/2$, $\pi/4$ and $0$, the length of the lines is proportional to the polarization of the radiation.}
\label{S_I}
\end{figure}

If the cavity contains relativistic electrons, the presence of the magnetic field will lead to a polarized synchrotron emission from the cavity. The intensity of the synchrotron radiation depends both on the magnetic field and the density of the electrons. We assume that the density of the electrons is related to the pressure we have evaluated by a relation of $\gamma=4/3$, as they are relativistic. The magnetic field is well defined by the solution we have found. Again we face the same uncertainties about the minimum pressure in the cavity, which we now consider to be equal to zero. We find that this structure produces synchrotron radiation. The apparent profile of the synchrotron radiation depends on the orientation of the cavity with respect to the observer, Fig.~(\ref{S_I}). 

We remark that there are observations which correlate low frequency radio emission with x-ray cavities \cite{Sch2005}. The intensity of synchrotron emission (see for instance \cite{LPG2005}), is given by 
\begin{eqnarray}
I_{\nu}=\frac{n+7/3}{n+1}\kappa(\nu)\int K_{e}B_{\perp}^{(n+1)/2} d l
\end{eqnarray}
where $n$ is the spectral index of the population of relativistic electrons, in our case we choose it to be $n=2$, $l$ is to be integrated along the line of sight, $B_{\perp}$ is the component of the magnetic field which lies perpendicular to the line of sight. The function $\kappa(\nu)$ is defined as $\kappa(\nu)=\frac{\sqrt 3}{4}\Gamma(\frac{3n-1}{12})\Gamma(\frac{3n+7}{12})\frac{e^{3}}{m_{e}c^{2}}\Big(\frac{3e}{2\pi m_{e}^{3}c^{5}}\Big)^{\frac{n-1}{2}}\nu^{-\frac{n-1}{2}}$, where $e$ is the electron charge, $m_{e}$ is the rest mass of the electron, $\nu$ is the frequency of the radio emission, and $\Gamma$ denotes the Euler gamma function.   $K_{e}$ can be found from this equation $n_{\epsilon}=K_{e}\epsilon^{-p}d\epsilon d\Omega$, where $n_{\epsilon}$ is the number density of the electrons having energy of $\epsilon$ to $\epsilon+d\epsilon$ and lie within solid angle $d\Omega$. Assuming that the maximum magnetic field in the cavity is $10\mu G$, its diameter is $10kpc$, its  angular size is $0.5 arcmin$ we find that the maximum intensity of the synchrotron emission at $100MHz$ shall be of the order of $\sim 5J$. This observations shall require resolution of the order of a few arcseconds in low frequencies, which is anticipated to be within the capabilities of LOFAR \citep{Rot2003}.

\subsection{Synthetic X-ray images}

\begin{figure}
\centering
\includegraphics[height=2.75cm]{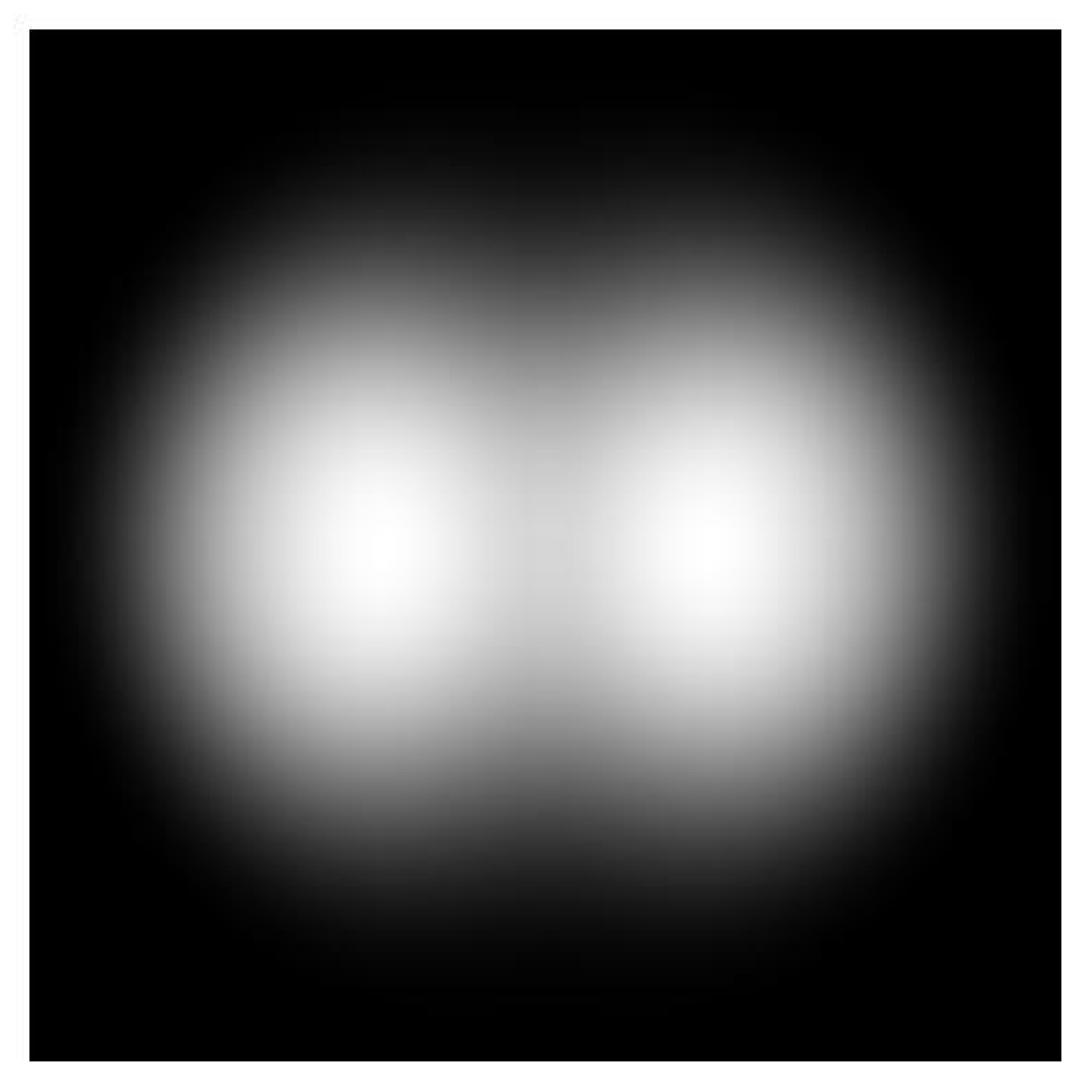}
\includegraphics[height=2.7cm]{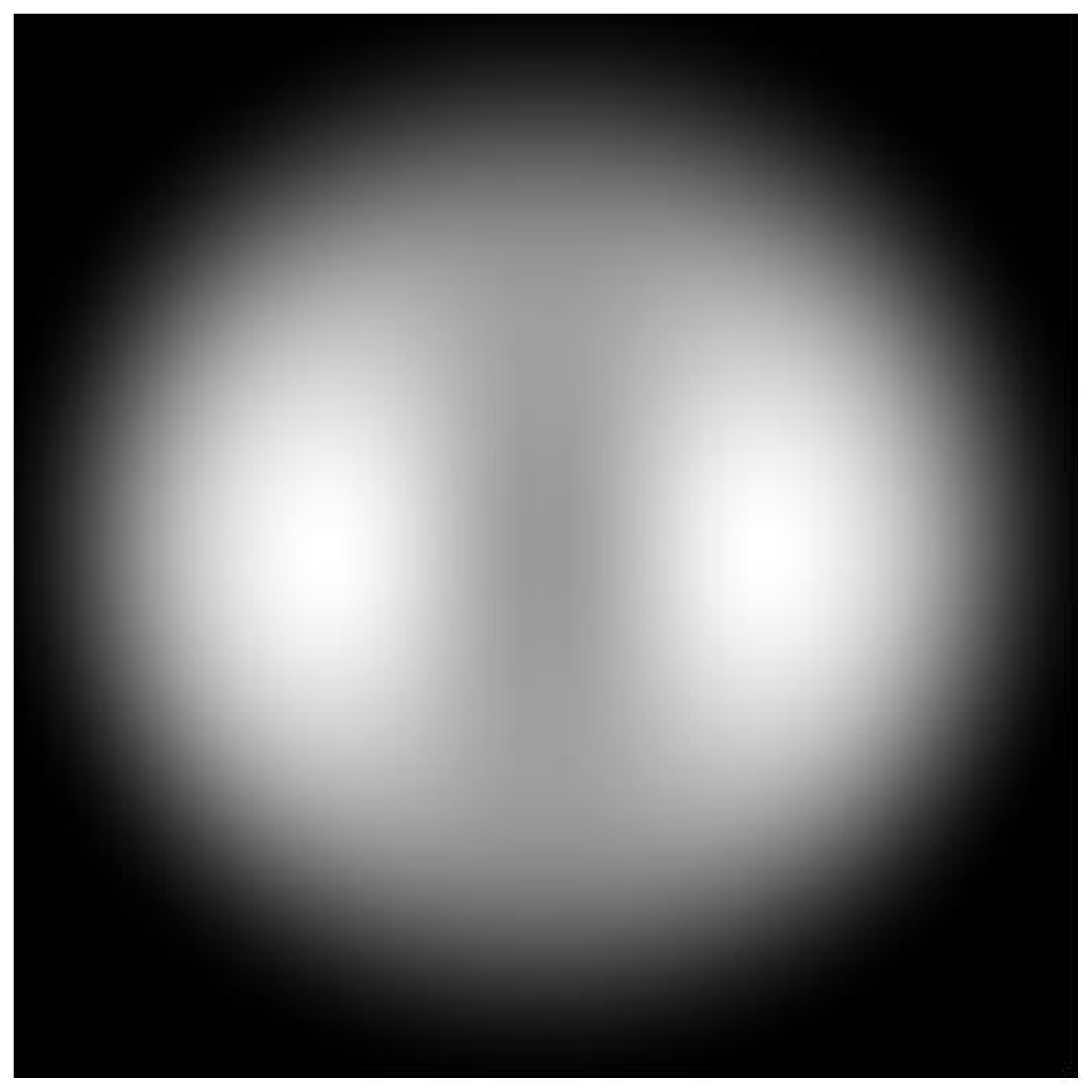}
\includegraphics[height=2.7cm]{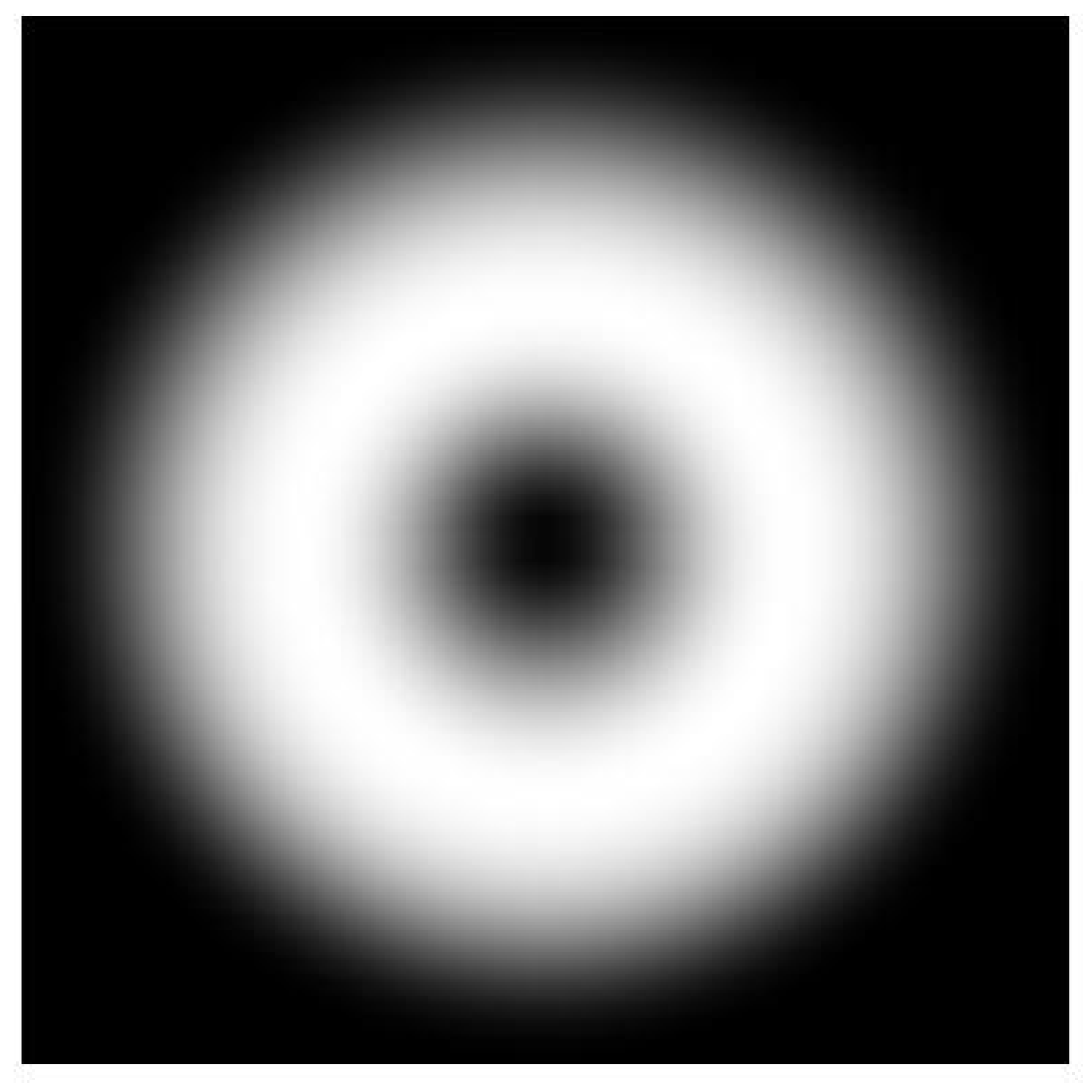}
\caption{Synthetic X-ray images for a magnetic cavity, the line of sight is normal to the axis in the first image, they form an angle of $\pi/4$ in the second one and they are parallel in the third. The white areas have lower X-ray brightness. The overall shape of the X-ray depression varies as the orientation changes, from a system with two lobes to an axially symmetric ring structure.}
\label{Xray}
\end{figure}

Observations of AGNs have revealed buoyant bubbles as depressions in the X-ray surface brightness. Following \cite{D2009} we evaluate the X-ray profile of the cavities. We assume that the X-ray emissivity of such a system is proportional to $E_{x} \sim \rho^{2} T^{1/2}$, applying an adiabatic relation with an index $\gamma=4/3$  as we have done in the rotation measure we find $E_{x} \sim p^{(3+\gamma)/(2 \gamma)}$. Then we integrate $E_{x}$ along the line of sight, taking into account the fact that the pressure in the external medium is constant and equal to $p_{0}$. Following this process we have constructed synthetic X-ray images for three different orientations so that the axis and line of sight have an angle of $\pi/2$, $\pi/4$ and $0$, Fig.~(\ref{Xray}). We remark that the shapes of the cavities vary from elliptical to spherical depending on the orientation, although their boundaries are always spherical. 

\section{Discussion}

The X-ray cavities appear as areas of lower X-ray emission in the intracluster medium, and they originate from AGN jets. Purely hydrodynamical simulations in general predict terminating shocks which are not observed in these systems, in addition they are vulnerable to instabilities and demonstrate ripples near their edges \citep{R2002}. MHD simulations \citep{N2006} using the context of magnetic towers \citep{L2003} suggest that the expansion of a magnetic jet in a background density and pressure predict the formation of collimated jet and a lobe in the top. \cite{D2009} have investigated models of buoyant bubbles for a variety of magnetic field configurations and viscosities, their simulations suggest that a toroidal field confined inside the cavity is the most promising, however the overall structure of the cavity changes drastically as it rises. 
 
In our study we do not discuss the process of the initial formation of the cavity but we focus on the equilibrium state after the inflation has taken place. A plausible physical description is the following: magnetic jets, after they have expanded, are dominated by toroidal field, which is now unstable. We suggest that through these instabilities the topology of the field changes and the cavities containing both toroidal and poloidal field are created.  

Then the cavities rise and expand in the intracluster medium, and there is a relation between the radius of the cavity and distance from the origin which is discussed in detail by \cite{D2008}. Our work describes a static cavity, a rising cavity will encounter a varying background pressure, however this does not pose an issue for our work, as it is feasible to balance it with an external environment by choosing appropriate values for $\alpha$ and $F_{0}$, provided that the expansion takes place slowly enough so that the system passes through states of equilibrium. However, a possible issue could be a pressure environment where the difference in background pressure is large enough over lengths of the size of the cavity, this may lead to deformation of the structure. 

In our model we have prevented the formation of surface currents, this was done to achieve a constant pressure on the surface. However if we had allowed the formation of surface currents following a force-free model, the cavity would have deformed, because the currents vary with the polar angle, thus the pressure in the inside of the cavity is not constant and cannot balance with a constant external pressure. 

 In addition we remark the difficulty of the simulations to treat surface currents, as they formally are a discontinuity which leads to tedious calculations. Observations are not conclusive whether there are surface currents or not, however the presence of surface currents has specific disadvantages we have discussed previously. We suggest that his solution can serve as an initial condition for simulations of rising bubbles. 

Coming to the issue of observable properties, they are consistent with the structures observed in X rays. However, X-ray observations do not give a conclusive answer on the presence of magnetic fields. Radio observations of polarized radio emission or  
  rotation measure of the system can give a more clear answer the question of the magnetic field. However, we are not aware of such measurement of sufficient resolution to determine the radio behaviour of the bubble. We remark however that magnetic fields are inferred   in large-scale structures coinciding with the Hercules and Perseus-Pisces superclusters \citep{X2006} and also a study of the polarization of Perseus cluster by \cite{dBB2005}.
  
 We remark that similar structures occur as ejecta in magnetar giant flares \citep{GKG2005}. They are created by strong magnetic field and plasma which is driven by the magnetic action on the magnetar, but they are scaled to smaller sizes. As our solution does not have an intrinsic scale can be applied to those configurations.

\section{Conclusions}

In this paper we have found analytical solutions for magnetic cavities without surface currents. The cavities contain a magnetic field with poloidal and toroidal components and a hot plasma. The structure of the fields is such so that they drop gradually to zero at the end of the cavity. Because of that, there are no surface currents, unlike the force-free fields which require surface currents. There is a strong qualitative similarity between this field configuration and that found with numerical methods by \citet{Braithwaite2010}. The non-force-free equilibrium found here must of course have a plasma-$\beta$ higher than about unity; a low-$\beta$ equilibrium must be force-free in the interior with a current sheet at the boundary. 

A numerical test has confirmed the stability of the simplest (fundamental radial mode) equilibrium. Furthermore, simulations have been used to examine the behavior of the equilibrium when the gas pressure drops by a large factor (via cooling) and takes the bubble from the $\beta\succeq 1$ regime into the $\beta\preceq 1$ regime. It is found that, as expected, the bubble becomes approximately force-free in the bulk with a thin layer of high current density at the boundary and along the axis of symmetry.

These structures correspond to rotation measures which are distinguishable from other structures and especially spheromaks. These are clear when the line of sight is normal to their axis, however they get saturated if the if the magnetic field is too weak. When the angle between the line of sight and the axis of the system is small they resemble spheromak structures more closely, but there are distinguishable properties visible with a sufficient resolution. In addition we have constructed synthetic X-ray images that show remarkable resemblance to x-ray cavities observed. 

 We remark that although a bubble has a spherical boundary it may appear as an ellipsoid in the X-ray observations, depending on the orientation. This accounts for the variety of the observed shapes. In any case the cavities may have intrinsically various shapes.  In addition we have evaluated the rotation measure that will affect a background source if they contain non-relativistic plasma and the synchrotron radiation and its polarization if they contain relativistic particles.

\bibliographystyle{mnras}
\bibliography{BibTex}

\end{document}